\documentclass[particles,article,accept,pdftex,moreauthors]{Definitions/mdpi}
\usepackage{xcolor}

\firstpage{1}
\pubvolume{1}
\issuenum{1}
\articlenumber{0}
\pubyear{2023}
\copyrightyear{2023}
\externaleditor{Academic Editor:  Peter Senger, Arkadiy Taranenko and Ilya Selyuzhenkov }
\datereceived{1 February 2023}
\daterevised{15 February 2023} 
\dateaccepted{22 February 2023}
\datepublished{ }
\hreflink{https://doi.org/} 

	\Title{$\Lambda$ and $\overline{\Lambda}$ Freeze-Out Distributions and Global Polarizations in Au+Au Collisions}
	
	\TitleCitation{$\Lambda$ and $\overline{\Lambda}$ Freeze-Out Distributions and Global Polarizations in Au+Au Collisions}
	
	\Author{ Nikita Tsegelnik $^{1}$\orcidN{}, Evgeni Kolomeitsev $^{1,2}$*\orcidA{} and Vadym Voronyuk $^{3}$\orcidV{}}
	
	\AuthorNames{Nikita Tsegelnik, Evgeni Kolomeitsev and Vadym Voronyuk}
	
	\AuthorCitation{Tsegelnik, N.; Kolomeitsev, E.;  Voronyuk, V.}
	
	\address{%
		$^{1}$ \quad Laboratory of Theoretical Physics, Joint Institute for Nuclear Research, RU-141980 Dubna, Russia; tsegelnik@jinr.ru\\
		$^{2}$ \quad Department of Physics, Faculty of Natural Sciences, Matej Bel University, 
		SK-97401 Banska Bystrica, Slovakia; 
		\\
		$^{3}$ \quad Laboratory of High Energy Physics, Joint Institute for Nuclear Research, RU-141980 Dubna, Russia; Vadim.Voronyuk@jinr.ru
	}
	\corres{Correspondence: evgeni.kolomeitsev@umb.sk}
	
	\abstract{The gold--gold collisions at $\sqrt{s_{NN}}=7.7$ and $11.5$\,GeV are simulated within the PHSD transport model.
		In each collision event, the spectator nucleons are separated and the fluidization procedure for the participants is performed.
		The local velocities are determined in the Landau frame and the kinematic and thermal vorticity fields are evaluated.
		We analyze the thermodynamic properties of the cells where $\Lambda$s and $\overline{\Lambda}$s were born or had their last interaction. Such cells contribute to the formation of the observed global polarization of hyperons induced by the thermal vorticity of the medium. The $\overline{\Lambda}$ polarization signal is found to be mainly determined by hot, dense, and highly vortical cells at the earlier stage of the collision, whereas the $\Lambda$ polarization signal is accumulated over the longer time and includes cells with lower vorticity.
		The calculated global polarizations for both $\Lambda$s and $\overline{\Lambda}$s agree well with the experimental finding by the STAR collaboration at energy $\sqrt{s_{NN}}=11.5$\,GeV. For collisions at $\sqrt{s_{NN}}=7.7$\,GeV, we can reproduce the STAR data for $\Lambda$ hyperons, but significantly underpredict the observed global polarization of $\overline{\Lambda}$.
		Furthermore, we consider the centrality dependence of the hyperon polarization in collisions at 7.7\,GeV. It increases with an increase of centrality, reaches a maximum at 65--75\% and then starts decreasing rapidly for peripheral collisions.}
	
	
	\def\om{\omega}

	\DeclareMathOperator{\rot}{rot}
	
	\newcommand{\lsim}{\stackrel{\scriptstyle <}{\phantom{}_{\sim}}}
	\newcommand{\gsim}{\stackrel{\scriptstyle >}{\phantom{}_{\sim}}}

\begin{document}
		
		\section{Introduction}\label{sec:intro}

		A definitive signal of the non-vanishing spin polarization of $\Lambda$ hyperons in heavy-ion collisions (HICs)~\cite{Adamczyk-Nature}, which put an end to the controversy of early experiments~\cite{Harris-Lb,Anikina} ignited the interest to theoretical studies of the evolution of spin degrees of freedom in dynamical strongly interacting systems. It also disproves earlier naive expectations~\cite{Panagiotou-86} that the polarization signal should cease with the transition to the quark-gluon matter since the Thomas-precession mechanism in the quark-recombination process~\cite{DeGrand-81} responsible for the polarization signal in p+p and p+A collision is not operative anymore.
		
		Several mechanisms leading to the polarization signal are discussed in the literature, cf. review~\cite{Becattini-Lisa-2020}. The statistical approach developed in~\cite{Becattini-Tinti2010,Becattini-Chandra2013,Fang-Pang-Wang2016,Becattini-Karpenko-Lisa2017} couples the local spin polarization of fermions with the local vorticity fields in the fireballs formed in {nucleus-nucleus collisions.}
		This mechanism implemented in hydrodynamic~\cite{Karpenko-Becattini2017,Xie-Wang-Csernai2017,Ivanov-PRC100,Ivanov-PRC102,Ivanov-PRC103,Ivanov-PRC105} and transport models~\cite{Li-Pang-Wang-Xia-PRC96,Sun-Ko-PRC96,KTV-PRC97,Wei-Deng-Huang-PRC99,Shi-Li-Liao-PLB788,Vitiuk-BZ2020}
		allows to reproduce the measured $\Lambda$ polarization. However, it is still difficult to explain the large $\Lambda$ polarization observed by the HADES collaboration~\cite{Yassine:2022} at low colliding energies $\sqrt{s_{NN}}\sim 2.5$\,GeV, {see Figure~3 in~\cite{Yassine:2022}.}
		Furthermore, most of the above works could not explain the larger polarization of $\overline{\Lambda}$s compared to $\Lambda$s if a special mechanism distinguishing particles from anti-particles is not included, see, e.g., \cite{Ivanov-PRC105,Ivanov-PRC102-AVE}.
		Remarkably, a splitting between $\Lambda$ and $\overline{\Lambda}$ polarization signals was successfully described within the UrQMD transport model~\cite{Vitiuk-BZ2020} where it was attributed to the different space-time distributions of $\Lambda$ and $\overline{\Lambda}$ and by different freeze-out conditions of both hyperons. {These calculations can also quantitatively describe the centrality dependence of $\Lambda$ polarization measured by HADES~\cite{Yassine:2022}.}
		
		In this paper, we continue our previous investigation~\cite{helicity:2022} of the vorticity and helicity fields in the hot and dense nuclear matter created in the HICs {at energies $\sqrt{s_{NN}}=7.7$\,GeV and 11.5\,GeV using the Parton-Hadron-String Dynamics (PHSD) transport model~\cite{PHSD,PHSD-contin}. These energies are reachable at the NICA facility and are within the STAR beam-energy scan range.}
		The goal is to investigate from which parts of the fireball the observed hyperons stem and at what conditions they are formed.
		Furthermore, we evaluate the global polarization of hyperons caused by the thermal vorticity field and study its centrality dependence.
		
		In Section~\ref{sec:fluid}, we review the results of our previous analysis of velocity and vorticity fields~\cite{helicity:2022}.
		In Section~\ref{sec:last-inter}, we discuss the distribution of thermodynamic conditions in the points, where the observed $\Lambda$ and $\overline{\Lambda}$ signals are formed. The resulting global polarization and its dependence on the collision centrality are considered in Section~\ref{sec:polar}. Conclusions are formulated in Section~\ref{sec:concl}.

		\section{Fluidization and Velocity and Vorticity Fields}\label{sec:fluid}

		The global hyperon polarization in HICs at collision energies $\sqrt{s_{NN}}=7.7$\,GeV was considered within the PHSD transport code
		in our previous work~\cite{KTV-PRC97}. In the recent work~\cite{helicity:2022} we reconsidered this problem making several essential improvements.
		
		So, at each time step we separate the spectator and participant nucleons, where the former ones are sorted out if their rapidity, $y$, is sufficiently close to the beam rapidity, $y_{\rm b}$, {namely $||y|-y_{\rm b}| \leq y_{\rm F}$, where $y_{\rm F}=0.27$ is the rapidity of the nucleon Fermi motion at the atomic nucleus density.} We assume that the spectators do not take part in the formation of the nuclear fluid, i.e., velocity and energy-density fields. The fluidization of the particle distribution generated by the transport code is performed in three steps. First, we define the cell grid and calculate the stress-energy tensor $T^{\mu \nu}$ and baryon current $J^\mu_B$ in each cell smearing the particles by the square-law spline function following the so-called cloud-in-cell method~\cite{Birdsall1997}.
		Then, the velocity 4-vector $u^{\mu}$ and the energy density $\varepsilon$ are defined as the eigenvector and eigenvalue of the energy-momentum tensor, correspondingly: $T^{\mu \nu}\, u_{\mu} = \varepsilon\, u^{\nu}$. The baryon density follows as the product $n_B=u_\mu J_B^\mu$. Knowing the energy density and the baryon density we can find the local temperature in the cell for the specific equation of state (EoS) as the solution of equation $\varepsilon=\varepsilon(n,T)$. As a model for the EoS, we use the hadron resonance gas at finite temperatures and baryon densities with density-dependent mean fields that guarantee the nuclear matter saturation~\cite{SDM09}. This EoS was used in hydrodynamic calculations~\cite{KT-HysHSD,KKT-Hydro}. Furthermore, calculating the kinematic and thermal vorticities we assume that only the grid cells with energy densities greater than the critical value $0.05$\,GeV$/$fm$^3$ are reliable for the thermal vorticity field formation. Note that this condition is not used as a condition of particle freeze-out. The evolution of each particle is traced till all interactions cease. Thereby, we avoid unphysically large gradients of the temperature occurring on the fireball boundary.

\begin{figure}[H]
			\includegraphics[width=0.3\linewidth]{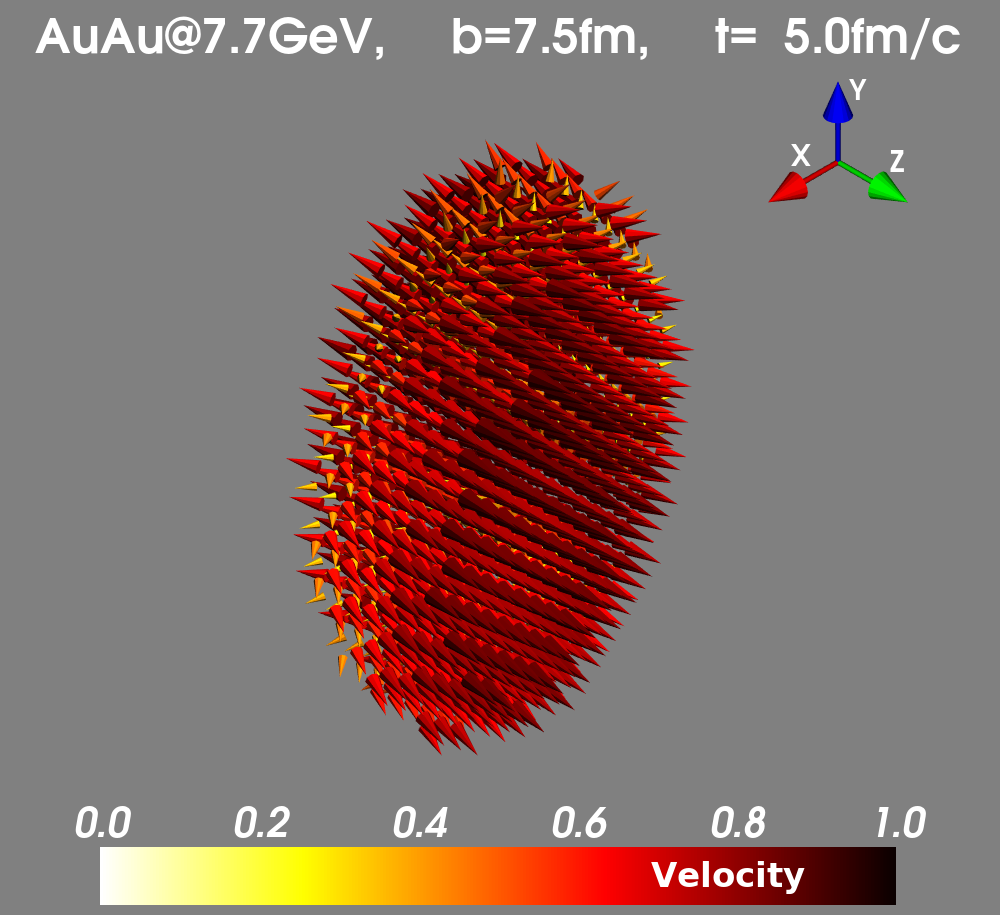}
			\includegraphics[width=0.3\linewidth]{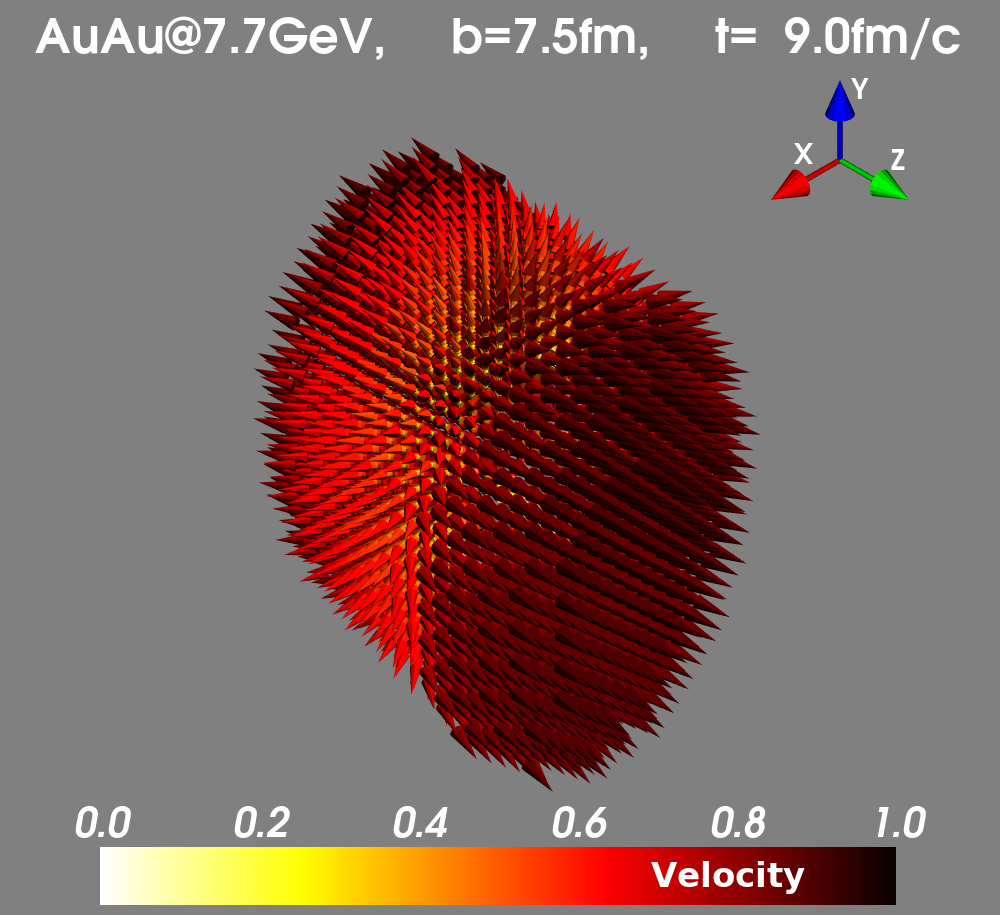}
			\includegraphics[width=0.3\linewidth]{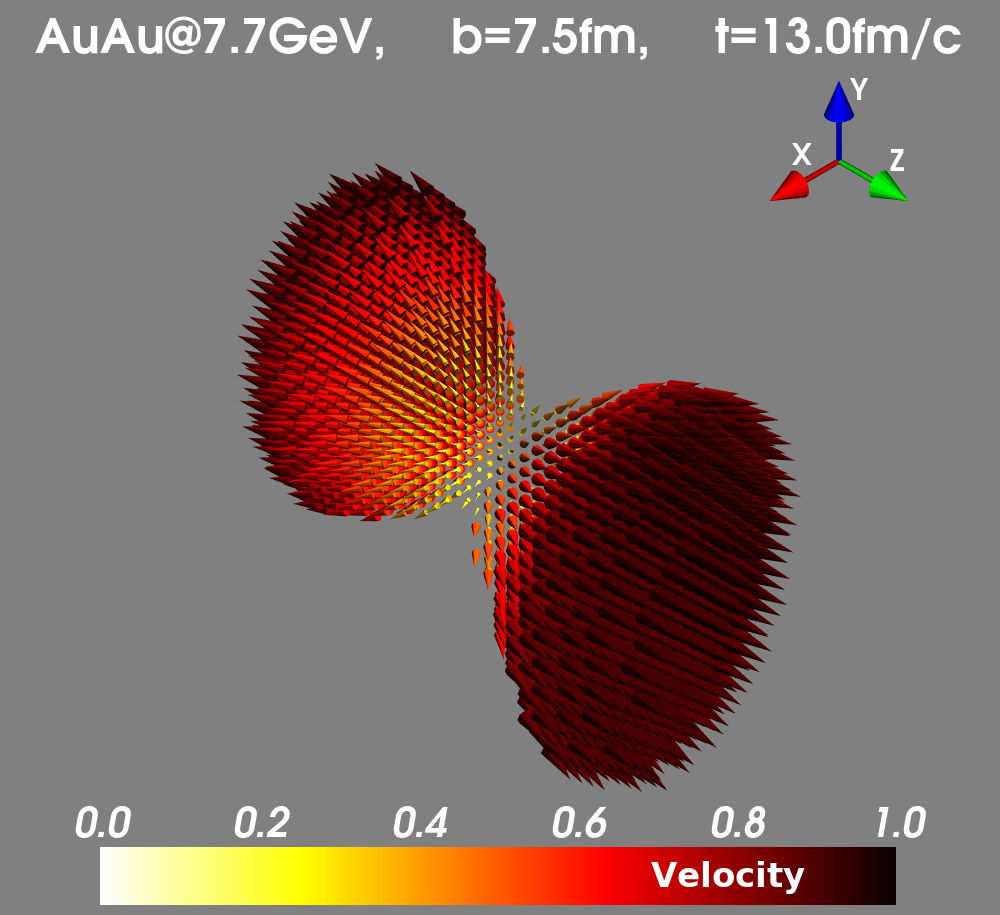}
			\caption{{The averaged fireball velocity fields created in Au+Au collisions at $\sqrt{s_{NN}}=7.7$\,GeV with the impact parameter $7.5$\,fm are shown for time moments $t=5,9$, and 13\,fm/$c$.}
				\label{fig:velocity}}
		\end{figure}

		Applying the above fluidization procedure, it was found in~\cite{helicity:2022} that the velocity field $\vec{v}$ resembles, to a large extent, the (2 + 1)D Hubble-like flow characterized by transverse and longitudinal velocities ($v_T$ and $v_\|$, respectively) with some small corrections of two types. The first type of correction is axially symmetric but introduces the weak dependence of the $v_T$ parameter on the longitudinal coordinate $z$ (the collision axis), and the $v_\|$ parameter on the transverse radius $r_T$. The second one is responsible for axial symmetry violation and, consequently, for a hydrodynamic elliptic flow. These additional terms lead to a non-vanishing kinematic vorticity $\vec{\omega} = \rot \vec{v}$ of the medium. The obtained velocity fields are illustrated in Figure~\ref{fig:velocity} for the Au+Au collision at $\sqrt{s_{NN}}=7.7$\,GeV with the impact parameter 7.5\,fm at times $t=5, 9$, and 13\,fm/$c$. The first moment corresponds (at given energy) to the maximum overlap (the hot zone is a disc with $|z|<2$\,fm) of colliding nuclei and the other times illustrate the fireball expansion. The total expansion time of the fireball till disintegration is about 15--16\,fm/$c$. The Hubble-like structure is clearly seen at $t=9$ and 13\,fm/$c$.
		
\begin{figure}[H]
			\includegraphics[width=0.3\linewidth]{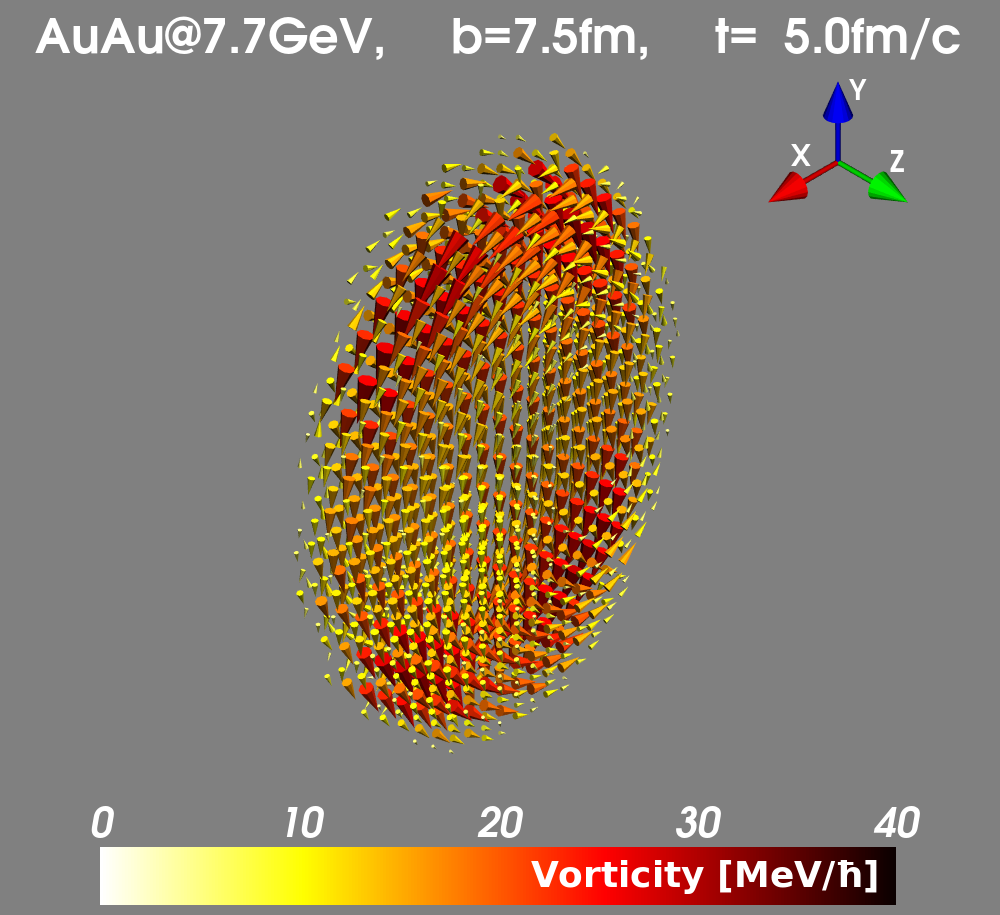}
			\includegraphics[width=0.3\linewidth]{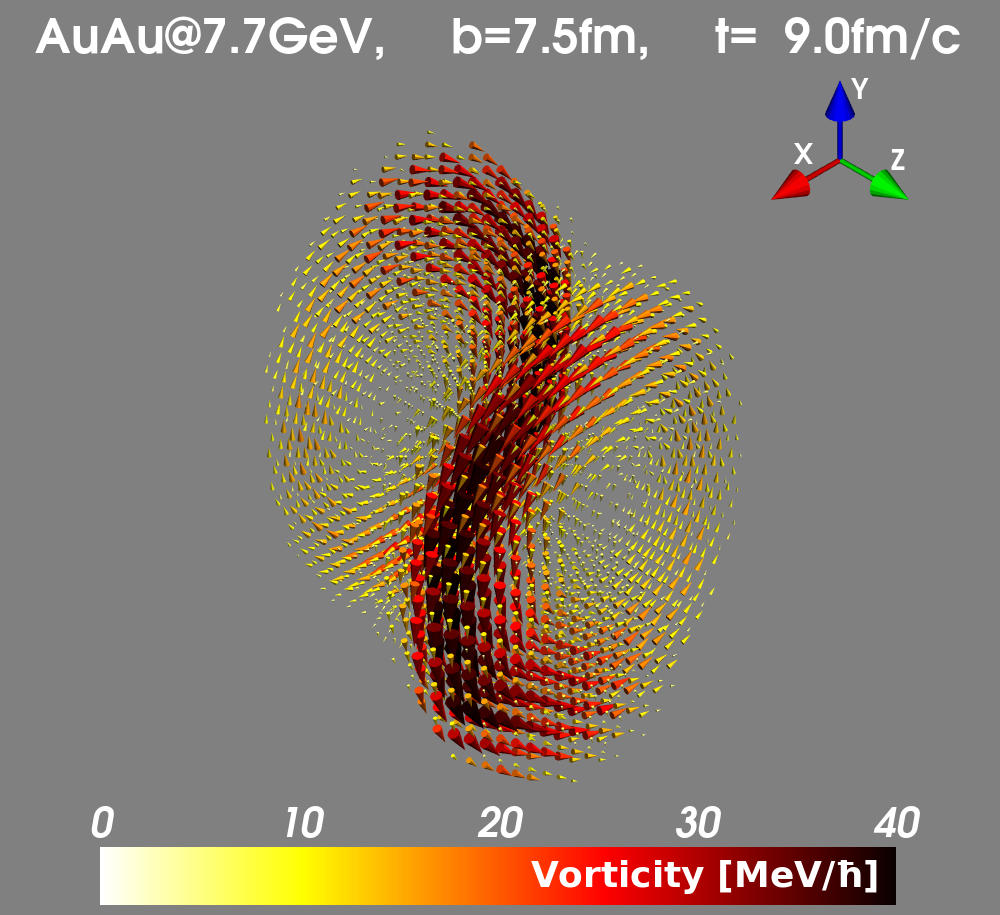}
			\includegraphics[width=0.3\linewidth]{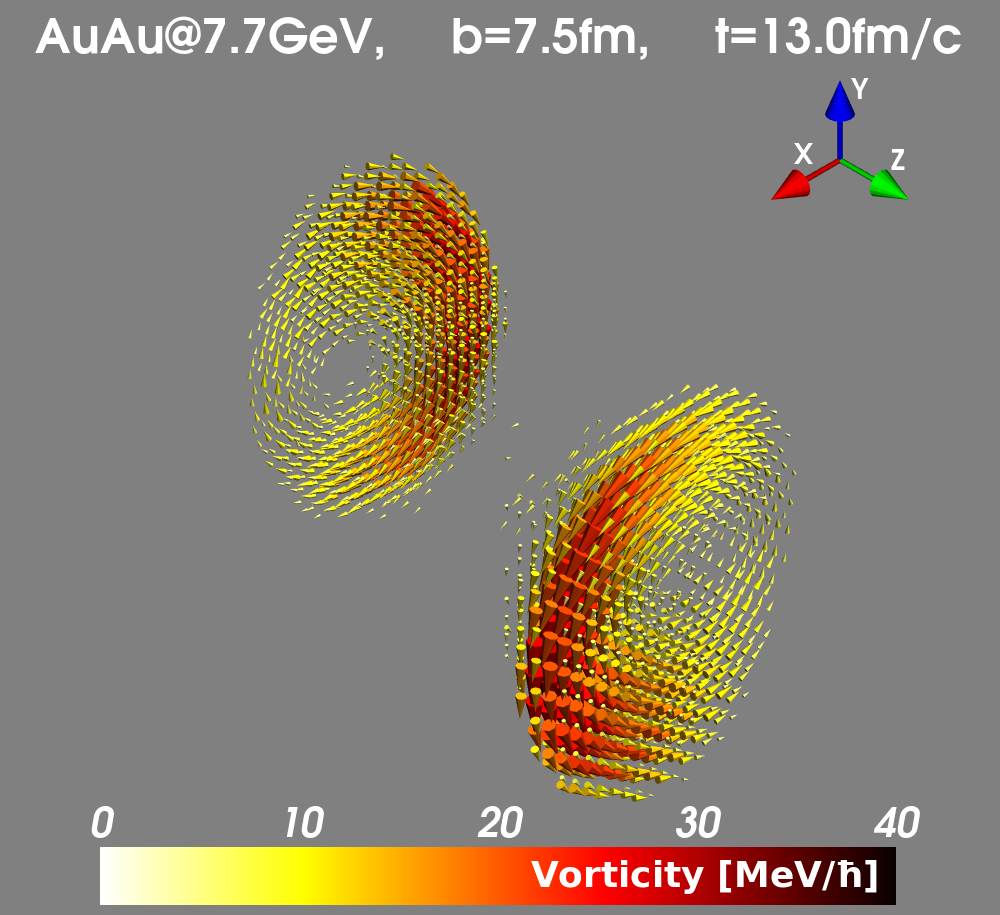}
			\caption{{The averaged fireball vorticity fields created in Au+Au collisions at $\sqrt{s_{NN}}=7.7$\,GeV with the impact parameter $7.5$\,fm are shown for time moments $t=5,9$, and 13\,fm/$c$.}
				\label{fig:vorticity}}
		\end{figure}

		The vorticity fields corresponding to the velocity distributions in Figure~\ref{fig:velocity} are presented in Figure~\ref{fig:vorticity}.
		At the maximum overlap time (5\,fm/$c$), the vorticity in the center slice is dominantly directed along the direction of the initial angular momentum of colliding nuclei (along the $y$ axis in our case). However, already at this time, a circular structure starts forming at the edges of the hot zone (the largest possible $z$). At later times one can recognize vorticity structures, which are called elliptic vortex rings.
		
		The irregular vorticity distribution and the distortions of the circular structure to the elliptic ones depend on the impact parameter. Uniform vortex rings are formed in central collisions, whereas in non-central collisions the elliptical structures are characterized by strong asymmetry in the $x-y$ plane. It is worth mentioning that maximum values of vorticity modulus reach $\hbar |\vec{\om}|\sim  60$\,MeV at $t\sim 7$\,fm$/c$.
		
		Similar vortex structures were first predicted at slightly higher energies in~\cite{Ivanov-rings} and confirmed in~\cite{Ivanov-rings-new} for the NICA energy range. Note, in our calculations, we apply the energy-density cut-off on the velocity and vorticity fields after their evaluations. Then, after averaging over collision events, the boundary layers of the system have vanishing vorticity. Otherwise, if the cut is applied before the derivatives are computed, numerical fluctuations and subsequent large gradients will lead to vorticity enhancement at the fireball boundary. Such behavior was observed in Refs.~\cite{BGST-Hseparation,BGST-Vsheet}, where a vortex sheet around the fireball was~obtained.

		\section{The Last Interaction Points of Hyperons}\label{sec:last-inter}
		
		The PHSD transport model and its predecessor---the HSD model~\cite{HSD}---proved to be successful in the description of strangeness production in HICs. In the current version, a special mechanism~\cite{Cassing:2015owa} is implemented, which increases the probability of the strangeness production in initial hard processes. The hard processes are described as the formation and breaking of a string in the framework of the FRITIOF Lund model~\cite{NILSSONALMQVIST1987387,Andersson:1992iq}. In the string decay, the quark-anti-quark pairs are assumed to be produced via the Schwinger mechanism \cite{Schwinger}. Then the relative probabilities of the production of pairs with different flavors are given by
		\begin{align}
			\label{eq:app:chiral:shwinger}
			\frac{P(s\bar{s})}{P(u\bar{u})} = \frac{P(s\bar{s})}{P(d\bar{d})} = \gamma_{s} = \operatorname{exp} \left( - \pi \frac{m^{2}_{s} - m^{2}_{u,d}}{2 \kappa}  \right),
		\end{align}
		where  $\kappa \approx 0.176$\,GeV$^{2}$ is the string tension coefficient and $ m_{u, d, s}$ are the masses of constituent quarks.
		For the FRITIOF's default settings with the constituent (dressed) quark masses $ m_u \approx 0.35 $\,GeV and $ m_s \approx 0.5 $\,GeV in the vacuum, Equation~(\ref{eq:app:chiral:shwinger}) gives the suppression of the strangeness production by a factor $ \gamma_s \approx 0.3 $ compared to the light-quark production. These relative production probabilities were corrected in Ref.~\cite{GCG-1998} for a better description of proton-nucleus collisions so that in the PHSD code the relative probabilities of the $u,d,s$ quark and $uu$ diquark production scale as
		\begin{align}\label{eq:app:chiral:production-ratio}
			u:d:s:uu = \begin{cases}
				1:1:0.3:0.007 \quad \mbox{for SPS and RHIC energies},\\
				1:1:0.4:0.007 \quad \mbox{for AGS energies}.
			\end{cases}
		\end{align}

		The main idea of Ref.~\cite{Cassing:2015owa} is to take into account in-medium modifications of the quark masses in the Schwinger's formula~\eqref{eq:app:chiral:shwinger} when the string breaking occurs in the dense and hot medium created in collisions.
		The reduction of the effective quark masses can be related to the  reduction of the quark condensate as
		\begin{align}
			\begin{split}\label{eq:app:chiral:mass-modif}
				m^{*}_{s} = m^{0}_{s} + (m^{V}_{s} - m^{0}_{s}) \frac{\langle \bar{q} q \rangle}{\langle \bar{q} q \rangle_{V}},\quad
				m^{*}_{q} = m^{0}_{q} + (m^{V}_{q} - m^{0}_{q}) \frac{\langle \bar{q} q \rangle}{\langle \bar{q} q \rangle_{V}},\quad q=u,d,
			\end{split}
		\end{align}
		with $ m^0_s \approx 100 $\,MeV and $ m^0_q \approx 7 $\,MeV for the `bare' quark masses.
		In Equation~(\ref{eq:app:chiral:mass-modif}), the effective masses of constituent quarks decrease from the vacuum values to the bare ones linearly with the reduction of the quark condensate. Substitution of these relations in the Schwinger's formula~(\ref{eq:app:chiral:shwinger}) implies the variation of the relative abundance of the strange quarks in dependence of the temperature and density of the surrounding medium. Within the PHSD code, the values of quark condensate are calculated in each cell on each time step~\cite{Cassing:2015owa}. These modifications of the string-breaking processes lead to significant improvements in the description of strange particle production at AGS and SPS energies~\cite{Palmese:2016rtq}.
		
\begin{figure}[H]
			
			\includegraphics[width=6.5cm]{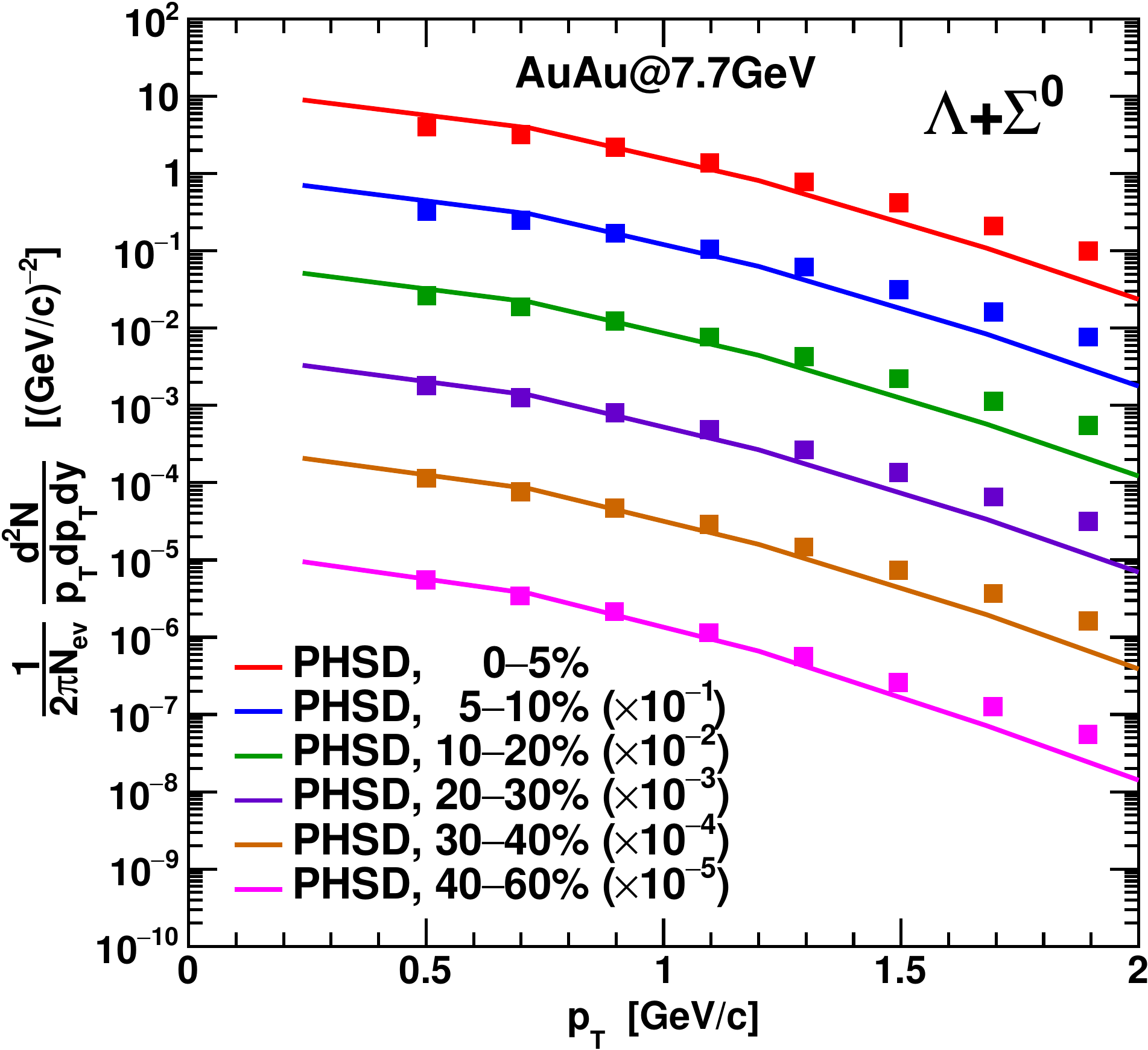}\,\,
			\includegraphics[width=6.5cm]{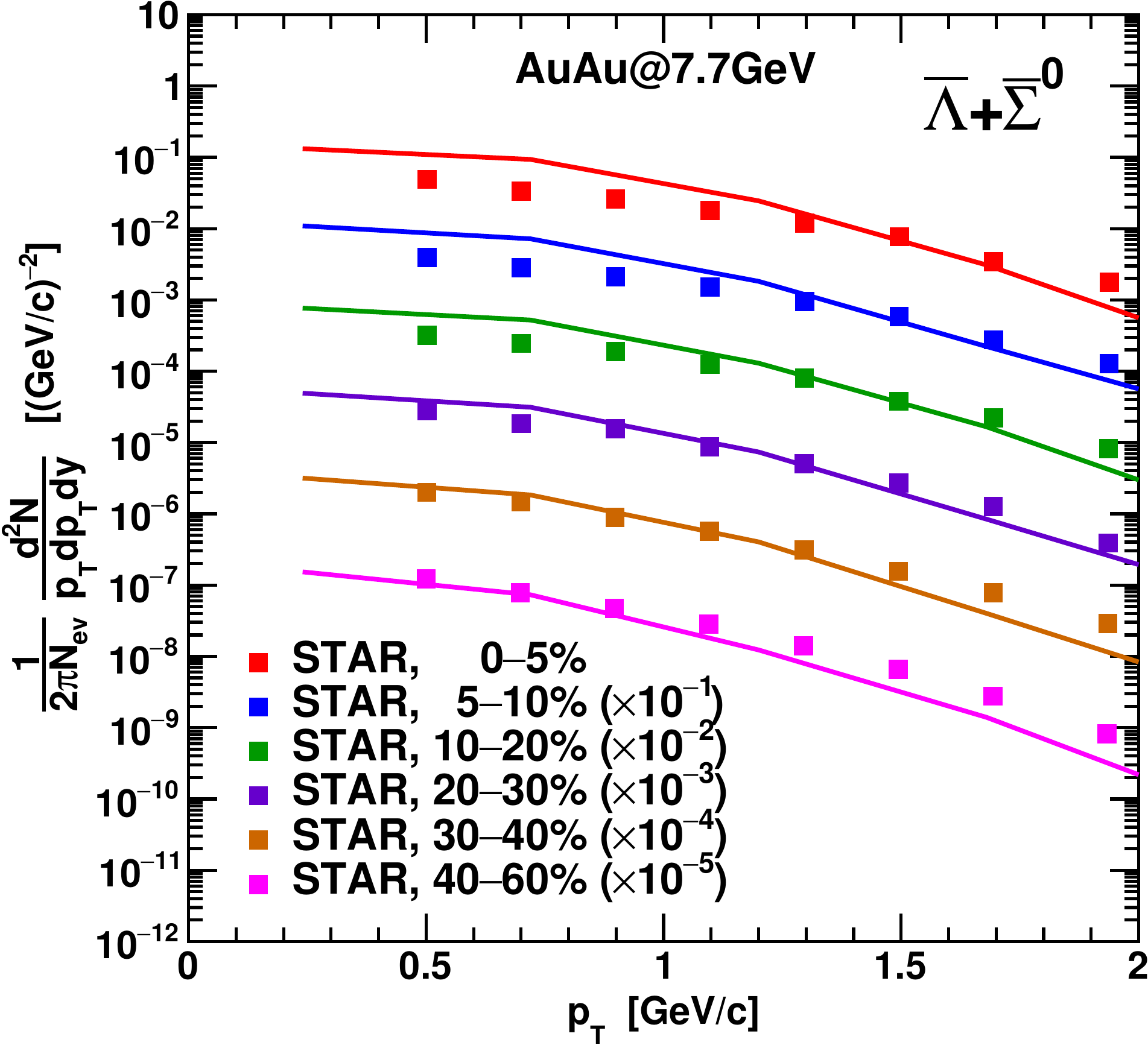}\\
			\includegraphics[width=6.5cm]{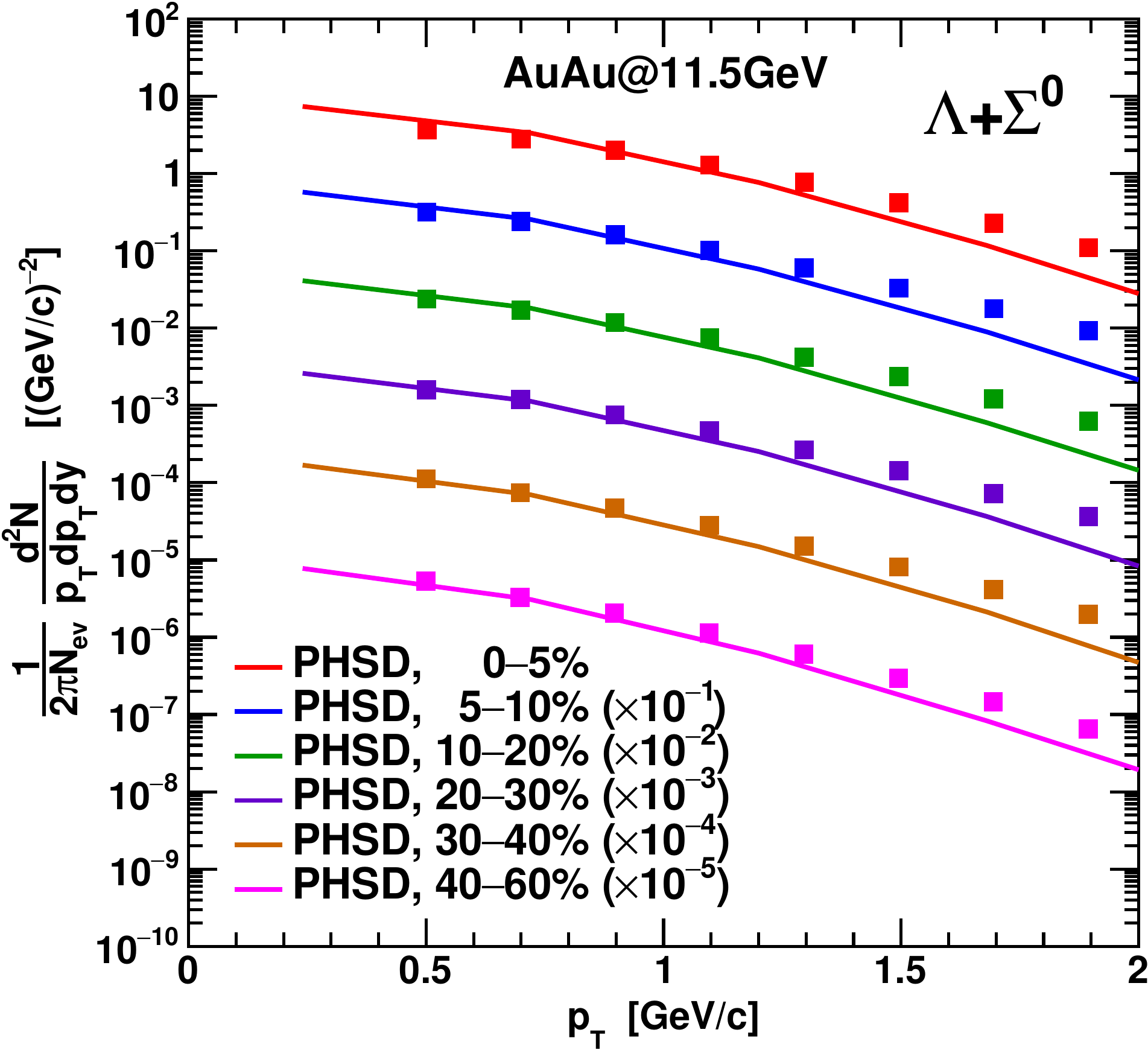}\,\,
			\includegraphics[width=6.5cm]{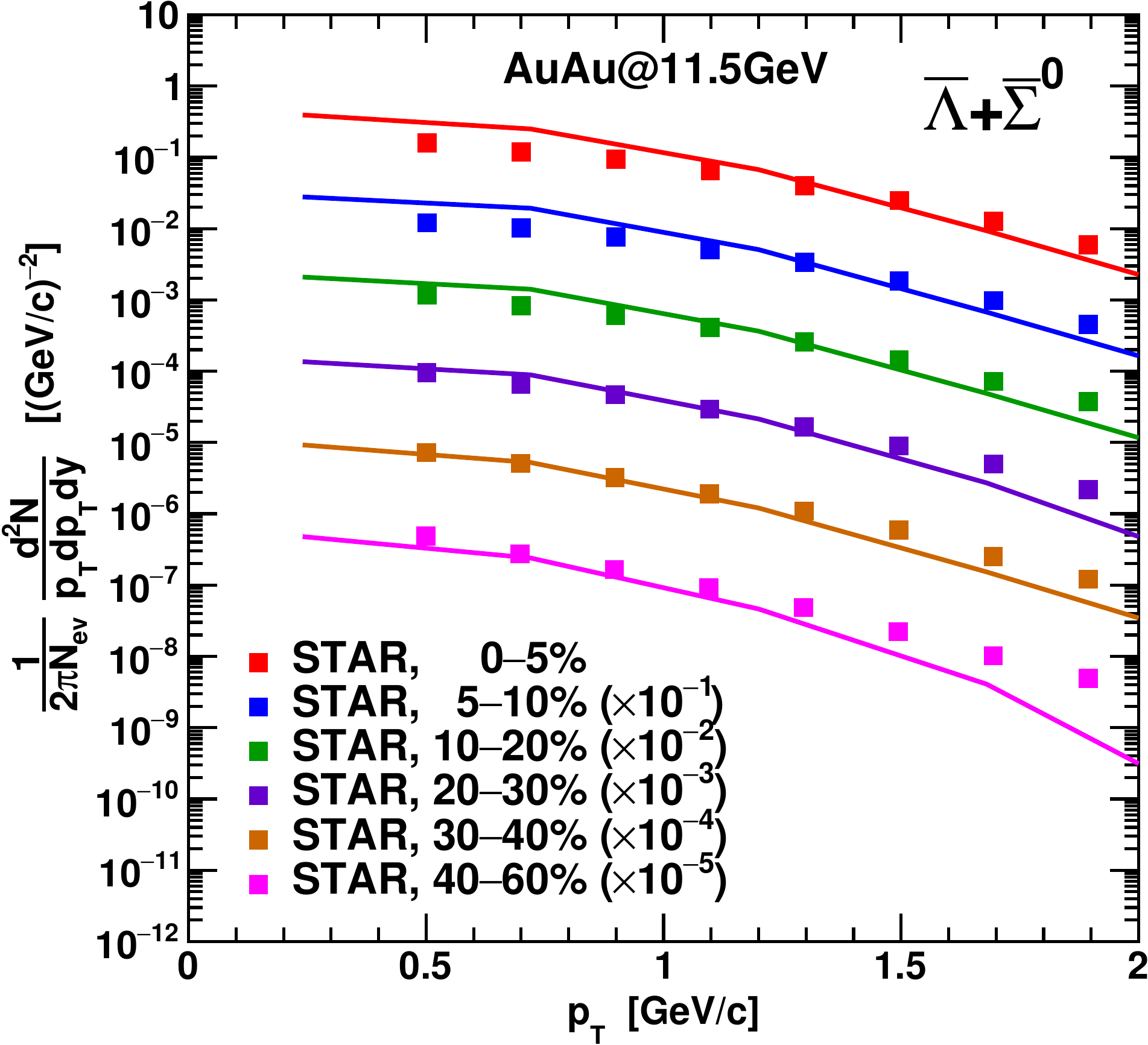}
			\caption{The transverse momentum, $p_T$, distributions of hyperons $(\Lambda+\Sigma^{0})$ (left column) and anti-hyperons $(\overline{\Lambda}+\overline{\Sigma}^{0})$ (right column) for Au+Au collisions at energies $\sqrt{s_{NN}}=7.7$\,GeV (upper row) and 11.5\,GeV (lower row) for the mid-rapidity $|y|<0.5$. Calculations correspond to various centrality classes. The curves obtained for centralities higher than 0--5\% are scaled down by the factors shown in Figure.  Dots show the experimental result obtained by the STAR collaboration~\cite{Adam:2019koz}.
				\label{fig:pt-spec}}
		\end{figure}

		In Figure~\ref{fig:pt-spec}, we illustrate the performance of the PHSD code in the description of the $\Lambda$ and $\overline{\Lambda}$ production
		at energies we consider in this article. The transverse momentum distributions of these particles at mid-rapidities are shown in comparison with the available experimental data obtained by the STAR collaboration. We see that the calculated $p_T$ distributions agree quite well with the experimental data for momenta $p_T<1.7$\,GeV. There is some deficit at higher momenta, which we expect could be corrected when we include the mean-field potential acting on particles between their collisions.

		{The main distinction of our current approach from the previous one in~\cite{KTV-PRC97} is the usage of dynamical freeze-out. During the code evolution, i.e., the evolution of the test-particle distributions in the coordinate-momentum space, we store additionally for each particle the information about the time of its last interaction (TLI) (including formation in the string breaking, resonance decays, and hadronic scatterings), the thermodynamic characteristics, velocity, and vorticity of the cell where it happens. At the end of the code run, we obtain distributions of these characteristics for each TLI. These distributions tell about properties of the medium, in which hyperons and anti-hyperons were frozen out, i.e., wherefrom they freely stream towards the detector without further interactions.
		}
		
\begin{figure}[H]
			
			\parbox{5cm}{\includegraphics[width=5cm]{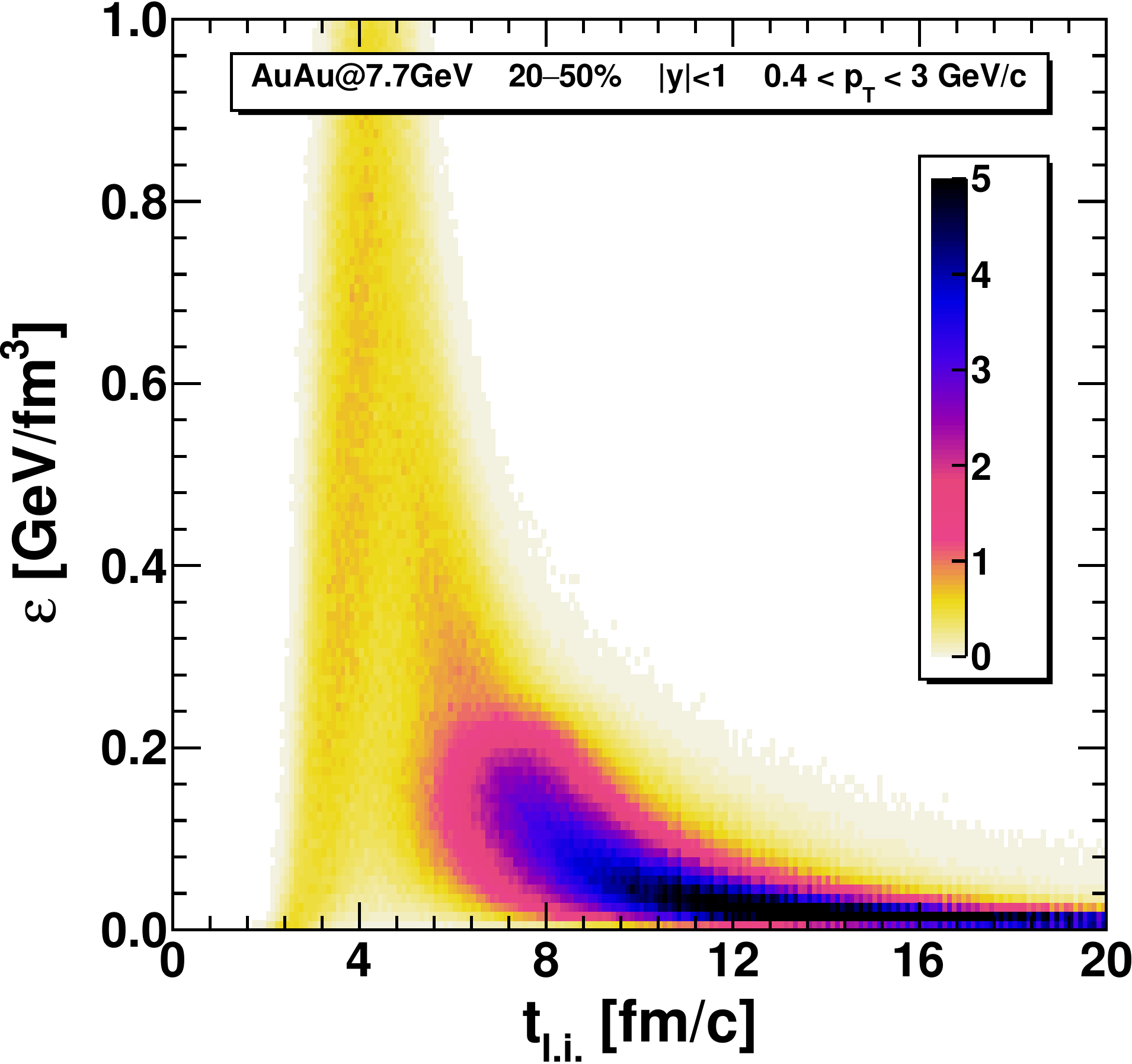}}\parbox{5cm}{\includegraphics[width=5cm]{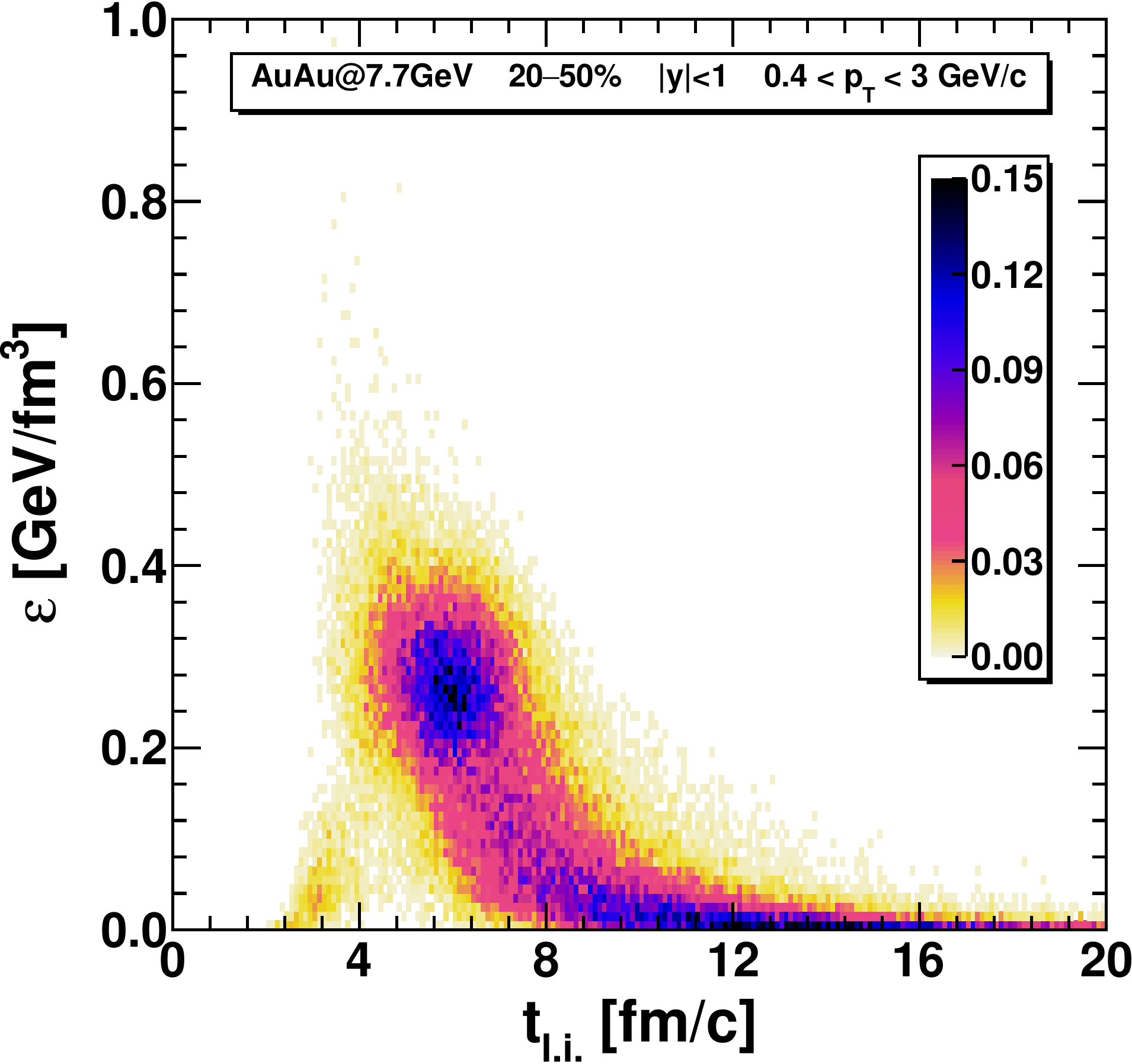}}\\
			\parbox{5cm}{\includegraphics[width=5cm]{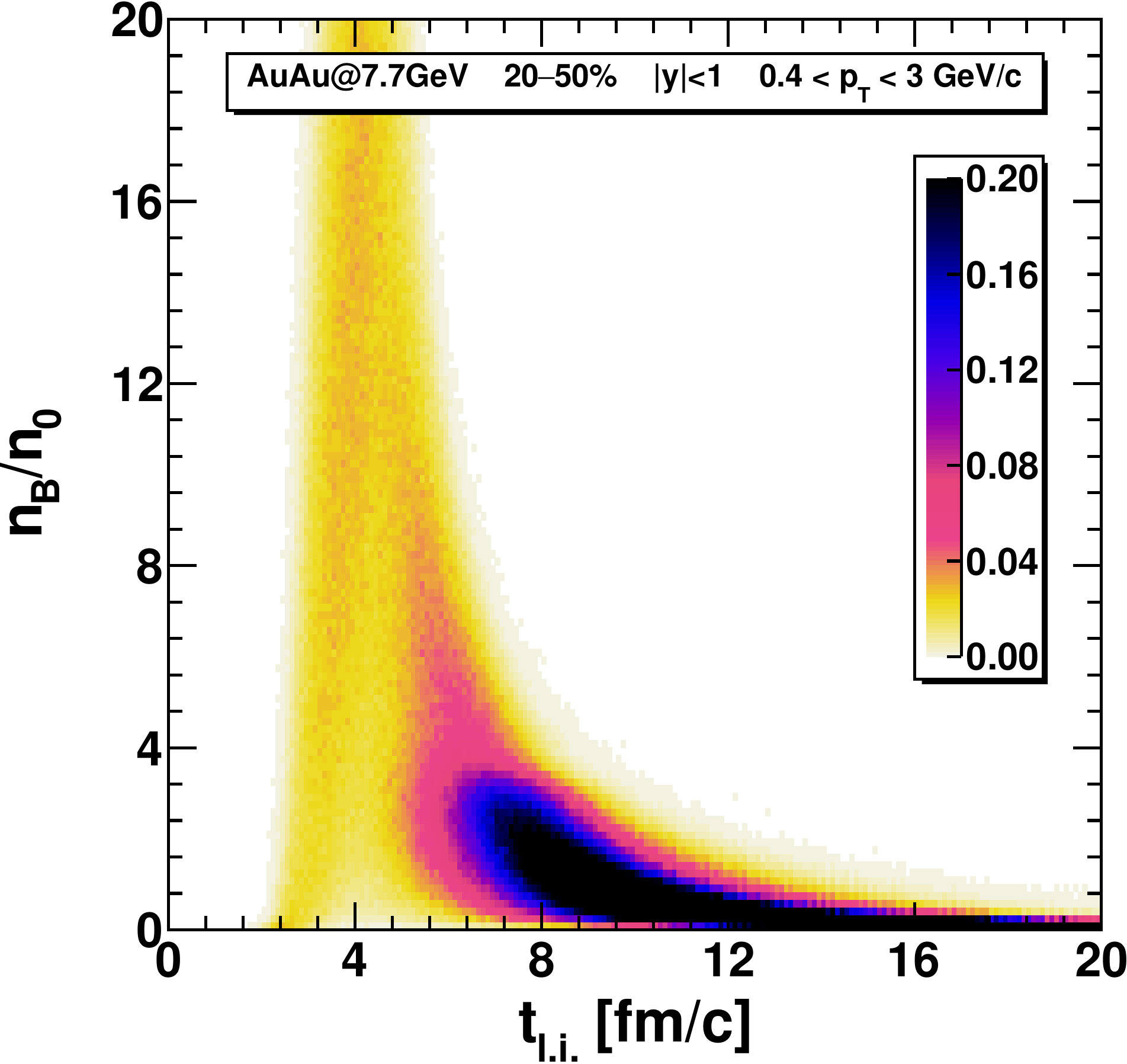}}\parbox{5cm}{\includegraphics[width=5cm]{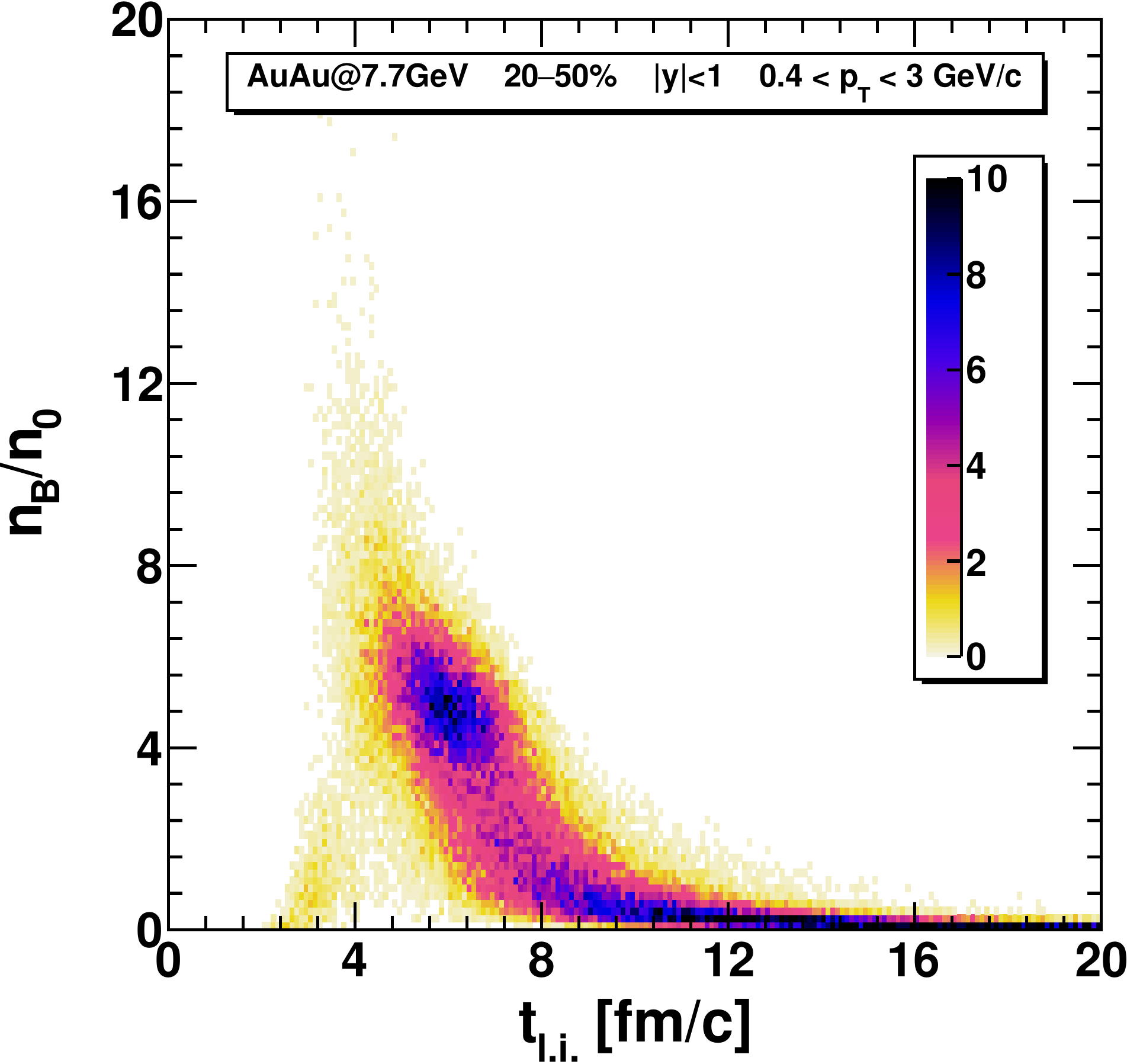}}\\
			\parbox{5cm}{\includegraphics[width=5cm]{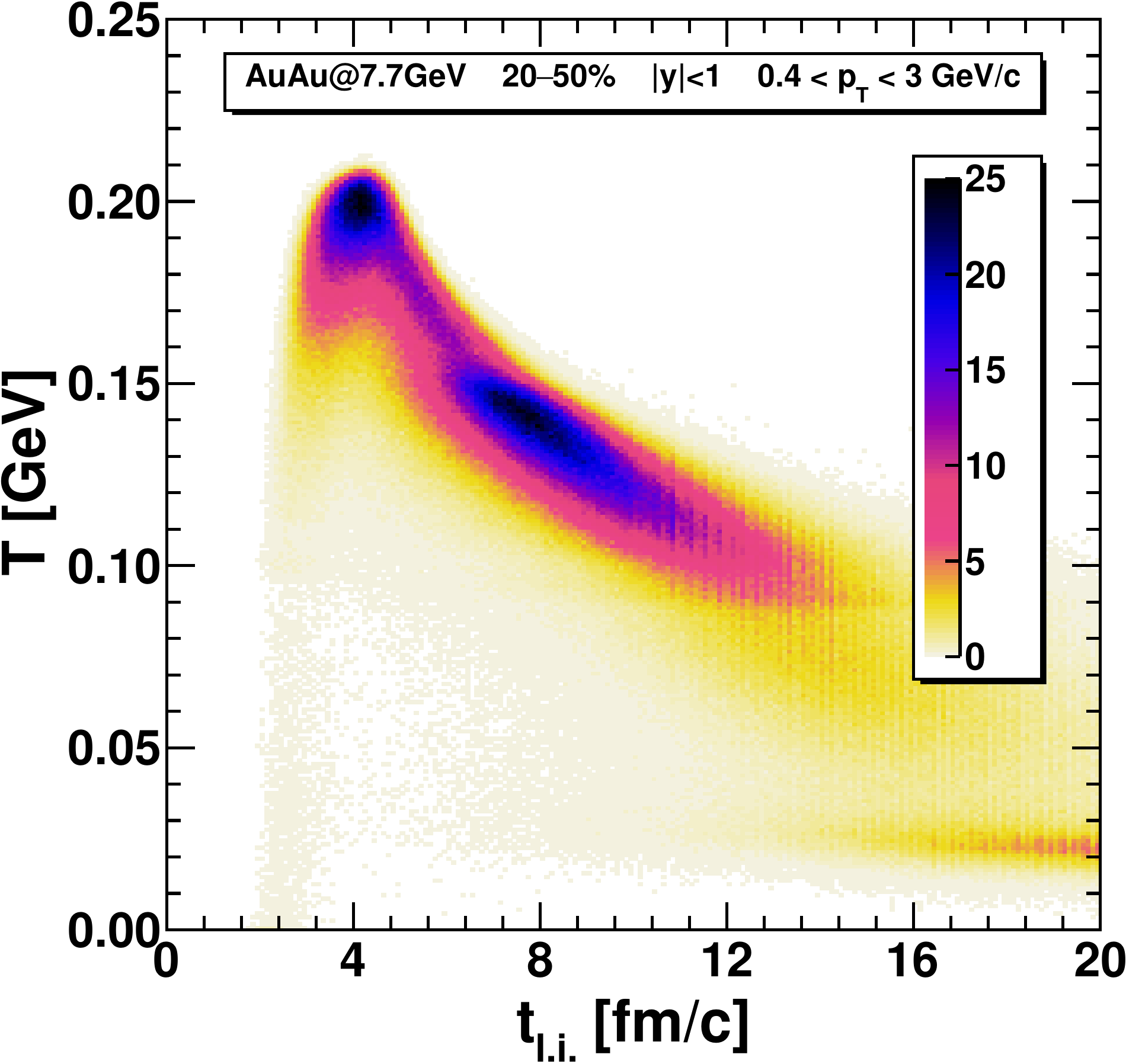}}\parbox{5cm}{\includegraphics[width=5cm]{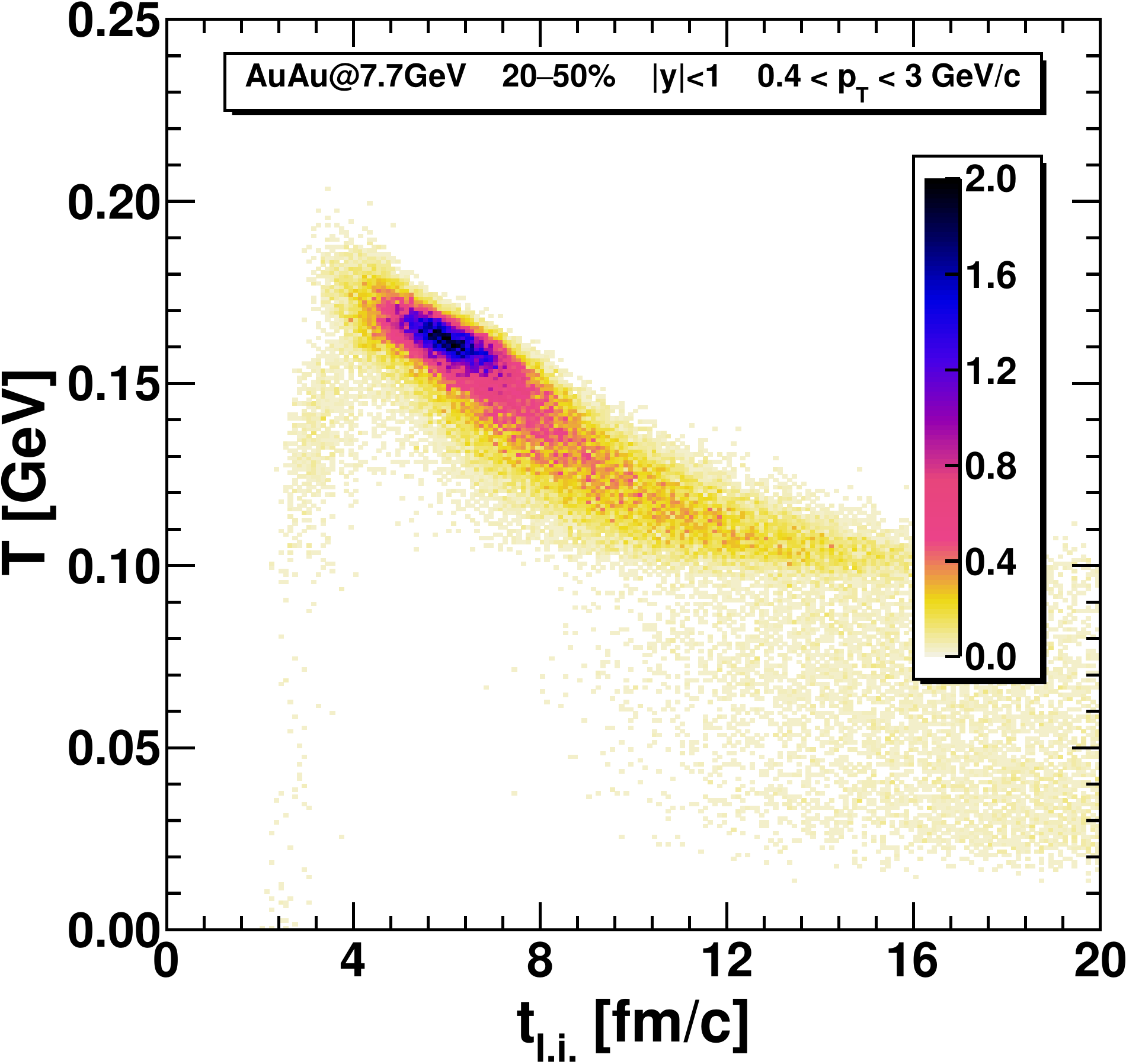}}\\
			\caption{The distributions at the TLI of the energy density $\varepsilon$ (first row), baryon density $n_B$ in units of the normal nuclear matter density $n_0=0.16$\,fm$^{-3}$ (second row), and temperature $T$ (third row) for final hyperons (left column) and anti-hyperons (right column) produced in the central 20--50\% Au+Au collisions at $\sqrt{s_{NN}}=7.7$\,GeV in the mid-rapidity range $|y| < 1$ for transverse momenta $0.4 < p_T < 3$\,GeV/$c$. All histograms are normalized on the number of collision events. The following bin width are used: 0.01\,GeV/fm$^3$ for the energy denstiy, $0.17\,n_0$ for the baryon density, 1\,MeV for the temperature, and 0.1\,fm/$c$ for $t_{\rm l.i.}$.}
			\label{fig:ENT-tli}
		\end{figure}

		The energy density, baryon density, and temperature distributions as functions of TLI, $t_{\rm l.i.}$, for hyperons (left column) and anti-hyperons (right column) are shown in \mbox{Figure~\ref{fig:ENT-tli}}. At earlier times ($2\,{\rm fm}/c\lsim t_{\rm l.i.} \lsim 4\,{\rm fm}/c$)
		the distributions of hyperons are very broad and spread to very high values of $\varepsilon\sim 1$\,GeV/fm$^3$, $n_B\sim20\,n_0$ and $T\sim 200$\,MeV. These distributions are due to the initial hard collisions (with string breaking) where the `unformed' hadrons are created and, then, propagate during the formation time. With some probability, they can propagate away from the fireball without further interaction after final formation. The second ``hot spot'' in the hyperon distributions, which is seen at times $7\,{\rm fm}/c\lsim t_{\rm l.i.}\lsim 13\,{\rm fm}/c$,
		corresponds to the hyperon production mainly in hadron-scattering processes and resonance decays.
		
		In contrast to hyperons, for the case of anti-hyperons, we observe only one significant source at $5\,{\rm fm}/c\lsim  t_{\rm l.i.}\lsim 7\,{\rm fm}/c$, and all the distributions are more localized. Note that the production of the most anti-hyperons corresponds to the maximum vorticity modulus and beginning of the vortex ring formation, see Figure~\ref{fig:vorticity}.

		Although a lot of hyperons stem from the hot and vortic regions formed at the early stages of collisions and carry, therefore, a high degree of spin polarization, roughly the same amount of hyperons is released at later stages from cells with smaller vorticities, reducing, thereby, total global polarization signal. For anti-hyperons, there is only one hot spot and one could anticipate that the anti-hyperon polarization could be somewhat higher.
		

		\section{Global Spin Polarization}\label{sec:polar}
		
		For the calculations of the hyperon spin polarization we use the thermodynamic approach~\cite{Becattini-Chandra2013,Becattini:2015ska}, where the local spin polarization of a particle with mass $m$ and spin $s$ is induced by local thermal vorticity:
		\begin{equation}\label{eq:becattini:thermal-vorticity}
			\varpi_{\mu\nu} = \frac{1}{2} (\partial_{\nu} \beta_{\mu} - \partial_{\mu} \beta_{\nu}),
		\end{equation}
		where $ \beta_{\nu} = u_{\nu}/T $, $u_\nu$---4-velocity, $T$---temperature. The hyperon spin vector $S^{\mu}$ in the leading order of $\varpi_{\mu\nu}$ is defined by the following expression:
		\begin{equation}\label{eq:becattini:S-def}
			S^{\mu}(x,p)=-\frac{s\, (s+1)}{6\, m}(1- n(x,p))\varepsilon^{\mu\nu\lambda\delta}\varpi_{\nu\lambda}p_\delta,
		\end{equation}
		where  $p^{\mu}$ is 4-momentum, $ n(x,p) $---distribution function. In our calculations, we assume the Boltzmann limit $(1- n(x,p)) \approx 1$. The hyperon polarization $\vec{P}=\vec{S^*}/s$ is determined by the spin vector recalculated in the rest frame of the hyperon:
		\begin{equation}\label{eq:becattini:S-boosted-vec}
			\vec{S^*} = \vec{S} - \frac{(\vec{S} \vec{p})\, \vec{p}}{E(E+m)},
		\end{equation}
		where we used $S_0 E=(\vec{S} \vec{p}) $ since $S^\mu p_\mu \equiv 0$.

		Within our dynamical freeze-out approach, if a particle is born inside the medium, i.e., the local energy density $\varepsilon> 0.05\,{\rm GeV/fm^3}$ from a non-decay process, the spin vector is calculated according to Equation~(\ref{eq:becattini:S-def}). If a particle was born outside the medium in some elastic or inelastic scattering process, the particle polarization is reset to zero. This means that these particles will be not polarized on average. In the PHSD model, strong hyperon decays are dynamically included in the evolution of the system. For strong decays $\Sigma^{*} \rightarrow \Lambda + \pi $ and $\Xi^{*} \rightarrow \Xi + \pi,$ only one-third of the polarization of the initial hyperon is transferred to the daughter hyperon~\cite{Becattini-Karpenko-Lisa2017}. Finally, the hyperons that have gone to infinity carry the information about the medium and polarization at their TLI.
		
		Applying this algorithm to hyperon or anti-hyperon species $H$ and averaging over all played-out collision events we can determine its averaged global polarization, which is the $y$ projection of the vector
		\begin{equation}
			\langle\vec{P}_{H}\rangle = 2\, \langle\vec{S^*}_{H}\rangle.
		\end{equation}
		The 
		$y$ axis is always normal to the reaction plane in our simulations. This is primary and not yet a final observable polarization signal, since the weak and electromagnetic decays, $\Xi \rightarrow \Lambda + \pi$ and $\Sigma^0 \rightarrow \Lambda + \gamma$,  are not taken into account.

\begin{figure}[H]
			
			\includegraphics[width=8cm]{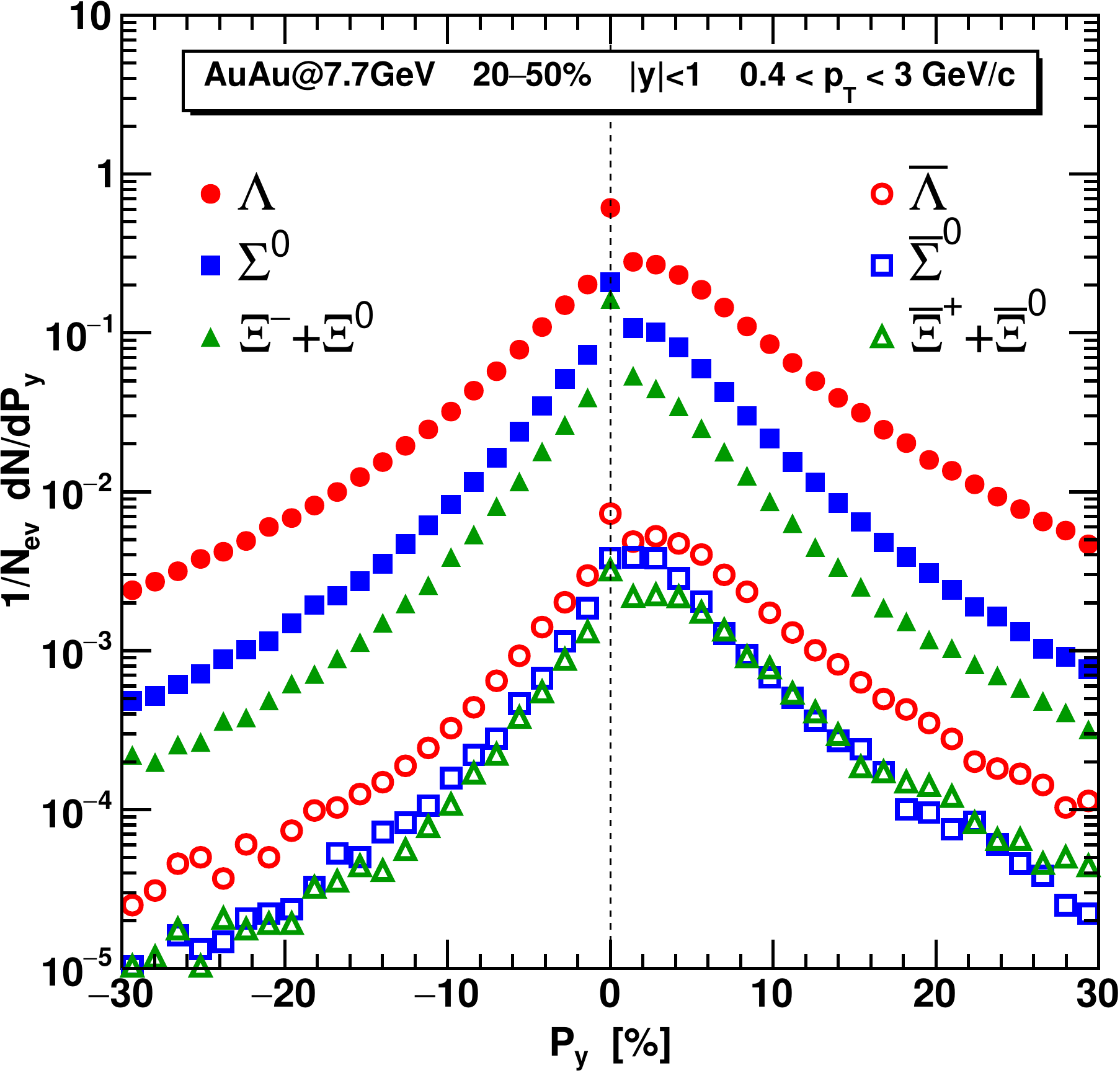}
			\caption{The  
				polarization distributions of the final hyperons and anti-hyperons (without weak and electromagnetic decays) in the semi-central 20--50\% Au+Au collisions at $\sqrt{s_{NN}}=7.7$\,GeV in the mid-rapidity $|y| < 1$ and transverse momenta $0.4\,{\rm GeV}/c < p_T < 3$\,GeV$/c$, normalized on the number of collision events.}
			\label{fig:dNdPolY}
		\end{figure}

		In Figure~\ref{fig:dNdPolY}, the polarization distributions for the final hyperons and anti-hyperons are shown. We observe the clear asymmetry between positive and negative polarization zones, so all the particles have non-vanishing averaged polarization.  One can see also the sharp maximum for $P_y=0$, due to the fact that for particles reborn outside of the medium, the polarization is put to zero. Therefore, there appears a discontinuity in the otherwise smooth distribution. From Figure~\ref{fig:dNdPolY} we conclude, that the number of secondary $\Lambda$s from $\Xi$ and $\Sigma^0$ decays is relatively large and taking into account the spin transfer coefficients, a significant reduction of the final polarization signal because of the feed-down can be expected.

		Applying the relations used Refs.~\cite{KTV-PRC97,Becattini-Karpenko-Lisa2017} to take into account the feed-down effects we obtain the final magnitudes for $\Lambda$ and $\overline{\Lambda}$ polarization, which are collected in Table~\ref{tab:PolY-sNN}.
		In our calculations, we observe strong suppression of the polarization due to the feed-down effect (up to $\approx36\%$). The main role in this effect is played by the decays of $\Sigma^0$($\overline{\Sigma}^0$) hyperons because of their large relative abundance.
		
		Nevertheless, we can reproduce the polarization signals for both $\Lambda$ and $\overline{\Lambda}$ at higher energy $\sqrt{s_{NN}}=11.5$\,GeV. The $\Lambda$ polarization for $\sqrt{s_{NN}}=7.7$\,GeV is also well described, however, the $\bar{\Lambda}$ polarization is significantly underestimated.
		
		\unskip
		\begin{table}[H]
			
			\caption{Global polarization of the $\Lambda$ and $\bar{\Lambda}$ hyperons in Au+Au collisions at $\sqrt{s_{NN}}=7.7$\,GeV and $\sqrt{s_{NN}}=11.5$\,GeV,  20--50\% centrality, mid-rapidity $|y|<1$ and transverse momentum range $0.4\,{\rm GeV}/c < p_T < 3\,{ \rm GeV}/c$. The STAR data~\cite{Adamczyk-Nature} is scaled to the currently accepted value of the decay parameter $\alpha_{\Lambda} = 0.732$~\cite{PDG:2020}.
				\label{tab:PolY-sNN}}
			\begin{tabularx}{\linewidth}{rrCCCC}
				\toprule
				&& \boldmath{$\Lambda$\textbf{(prim.)}} & \boldmath{$\Lambda$} & \boldmath{$\bar{\Lambda}$}\textbf{(prim.)} & \boldmath{$\bar{\Lambda}$}\\
				\midrule
				\raisebox{-3mm}{$\sqrt{s_{NN}}=7.7$\,GeV} & PHSD & 2.14 & 1.45 & 3.63 & 2.31\\
				&STAR &  & $1.79\pm0.58$ & & $7.60\pm3.25$\\
				\midrule
				\raisebox{-3mm}{$\sqrt{s_{NN}}=11.5$\,GeV} &PHSD & 1.63 & 1.12 & 2.64 & 1.77 \\
				&STAR &  &$1.18\pm0.39$ & &$1.58\pm 1.11$ \\
				\bottomrule
			\end{tabularx}
		\end{table}

\begin{figure}[H]
			
			\includegraphics[width=6cm]{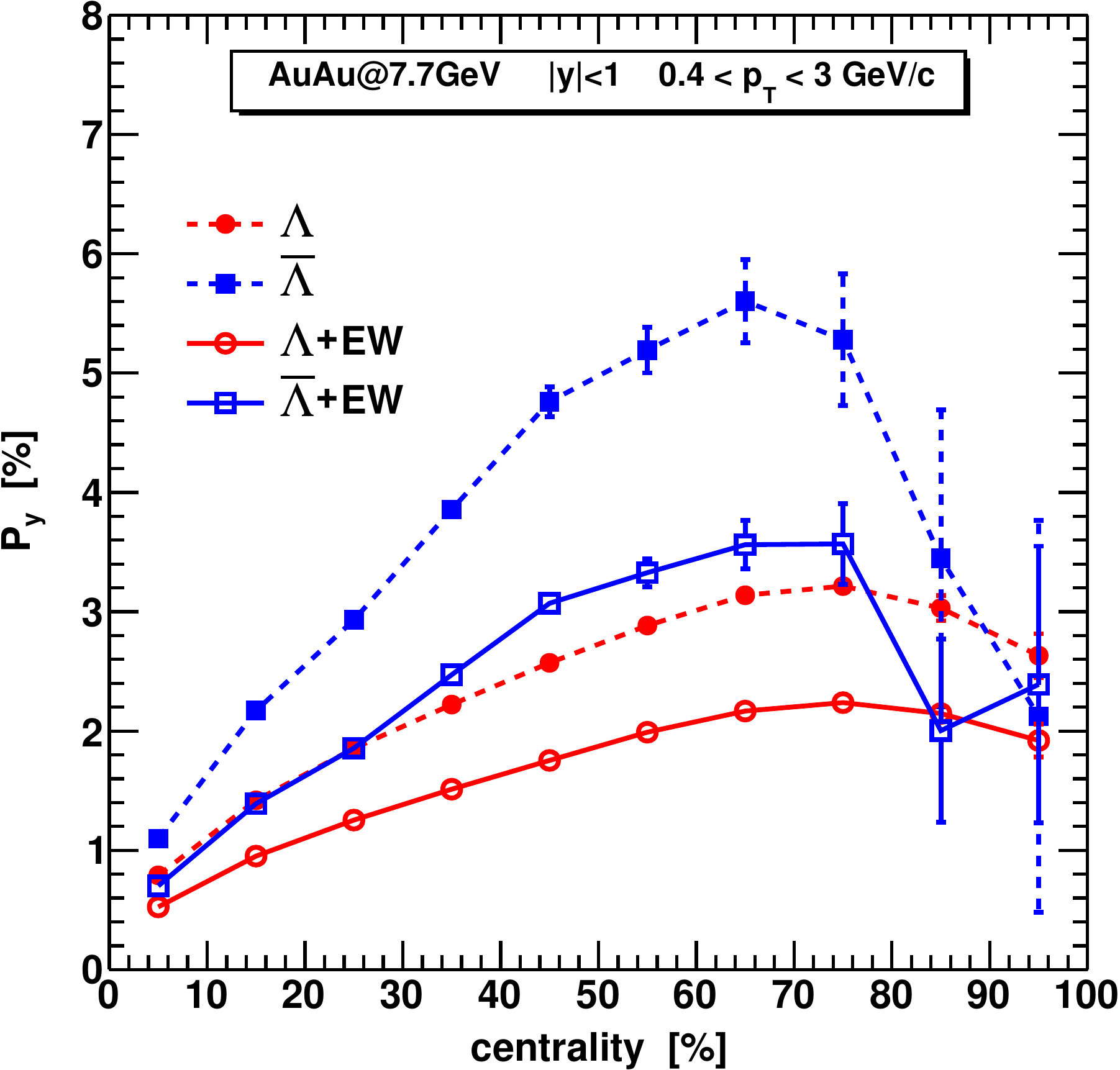}\quad \includegraphics[width=6cm]{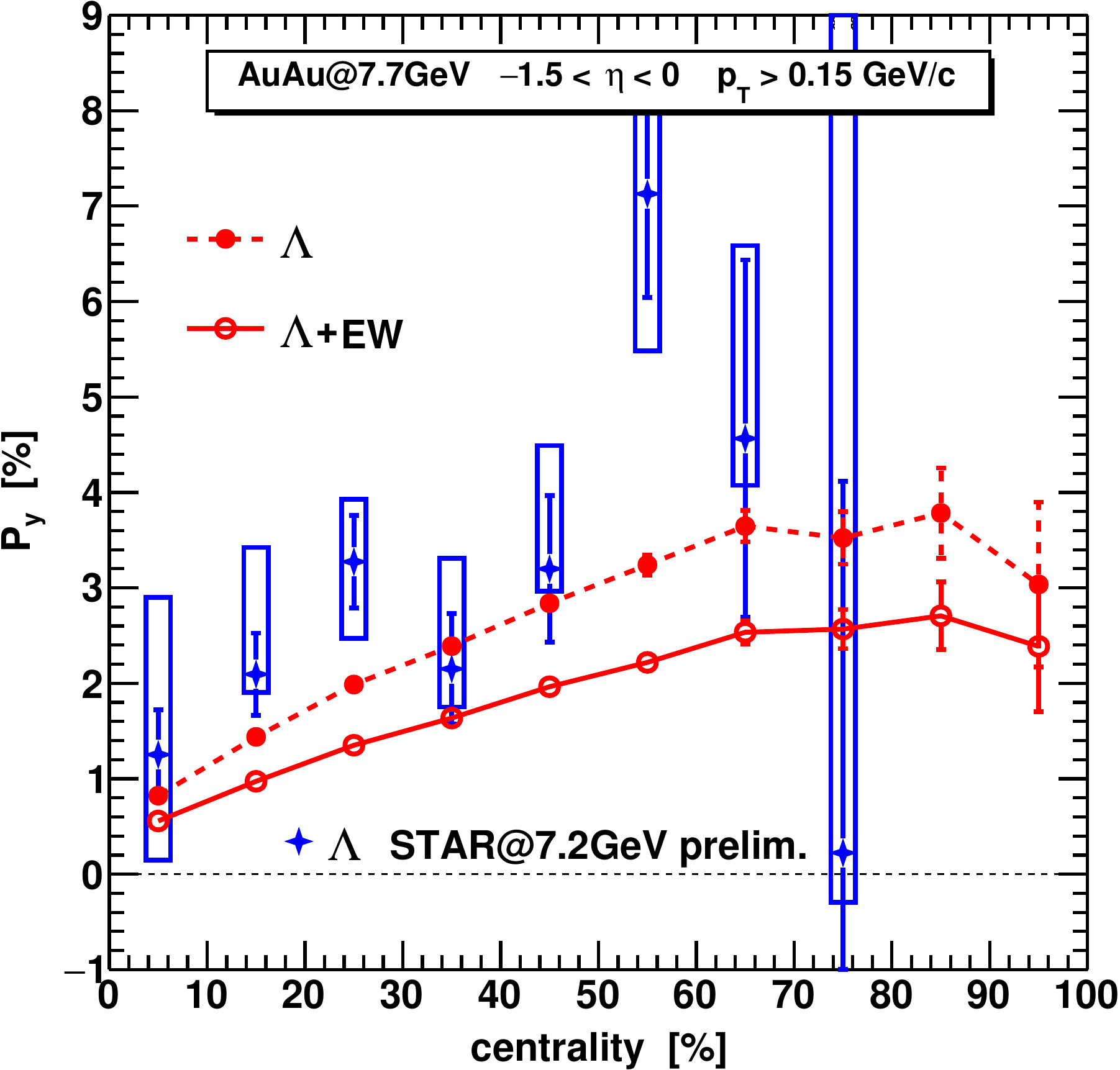}
			\caption{{Global 
					polarization  of $\Lambda$ and $\bar{\Lambda}$ depending on the collision centrality in Au+Au collisions at $\sqrt{s_{NN}}=7.7$\,GeV is shown for the mid-rapidity $|y| < 1$ and transverse momentum $0.4 < p_T < 3$\,GeV$/c$ regions on the left panel and
					for the pseudo-rapidity range $-1.5<\eta<0$ and $p_T>0.15$\,GeV/$c$ on the right panel. Star symbols on the right panel show the results of the STAR collaboration~\cite{Okubo:2022} for the same $p_T$ and rapidity ranges but energies $\sqrt{s_{NN}}=7.2$\,GeV.}}
			\label{fig:PolY-centr}
		\end{figure}

		In Figure~\ref{fig:PolY-centr}, (left panel) we present the dependence of the polarization signal on the centrality of the collision, $C$. Our calculations demonstrate the tendency of the polarization to increase with the increasing centrality for $C\lsim 65\mbox{--}75\%$ and to decrease for larger values of $C$.
		We associate this pattern with the proper separation of spectator and participant nucleons done in our calculations.
		A similar pattern is seen in preliminary data of the Au+Au collisions at 7.2\,GeV reported by the STAR collaboration in~\cite{Okubo:2022}, although for quite different $p_T$ and rapidity cuts. {On the right panel in Figure~\ref{fig:PolY-centr} we present the direct comparison of these data with the results of our calculation at 7.7\,GeV but with the experimental cuts. The magnitude of the observed polarization signal is somewhat larger than that in our calculations, however, the rising trend is clearly visible. The drop of polarization is more rapid in the data and starts at smaller centralities.}
		At lower collision energies, the continuous increase of the polarization with centrality for $C\lsim 50\%$ is seen in Au+Au collisions at $\sqrt{s_{NN}}=3$\,GeV in~\cite{STAR:2021beb}. Furthermore, in collisions at $\sqrt{s_{NN}}=200$\,GeV the linear rise of the polarization with $C$ for $C<60\%$ and further leveling off at larger $C$ was seen in the data in Ref.~\cite{Adam:2018}.

		Remarkably, other theoretical models~\cite{Ivanov-PRC105, Liang_2021, Ayala:2022yyx, Alzhrani:2022dpi}, which give a good description of the global polarization on the collision energy, do not produce a drop of the polarization for peripheral collisions. The geometrical model~\cite{Liang_2021} at $\sqrt{s_{NN}}=200$\,GeV predicts a plateau at $80\%$ centrality. The core-corona model~\cite{Ayala:2022yyx} at $\sqrt{s_{NN}}=3$\,GeV predicts a sharp drop of the polarization already for the 20--40\% centrality class. The hydrodynamic models~\cite{Ivanov-PRC105,Alzhrani:2022dpi} at different energies show a permanent increase of the polarization up to the largest values of~$C$.
		
		In Figure~\ref{fig:PolY-centr}, we see that the maximum of the global polarization for $\Lambda$ is reached for the $70\mbox{--}80\%$ centrality region, while for $\bar{\Lambda}$ the maximum is realized for $C=60\mbox{--}70\%$.  It is worth mentioning, that the angular momentum transferred to the medium for this class is relatively small compared to the $0\mbox{--}20\%$ and even $20\mbox{--}40\%$, see Figure~4 in~\cite{helicity:2022}. Therefore, the polarization cannot be explained by the angular momentum transfer only.

		\section{Conclusions}\label{sec:concl}
		
		We performed fluidization of the particle distributions generated by the PHSD transport code for gold--gold collisions at $\sqrt{s_{NN}}=7.7$ and 11.5\,GeV and determined evolutions of local energy density, baryon density, and temperature. Furthermore, the velocity and vorticity fields are calculated assuming that the velocity is defined within the Landau frame. Spin polarization of hyperons and anti-hyperons induced by the local thermal vorticity are evaluated. For each (anti-)hyperon leaving the fireball and ceasing participation in strong interactions we trace back its history to the moment its last interaction (TLI) and determine thermodynamical characteristics and vorticities of the regions where the (anti-)hyperons came from. The analysis of the TLI distributions for the energy density and the temperature reveals that there are two sources of final hyperons: one at the earlier stage of the collision due to a hard process in hot cells with large vorticities and the second one at later times when final hyperons stem mostly from resonance decays and hadron scatterings in cooler and less vortical cells. In contrast, final anti-hyperons have only one source at an early stage of the collision. Thus, the hyperon polarization signal acquired at the earlier stage is diminished by contributions from the later stages relative to the anti-hyperon polarization signal. The final polarization of $\Lambda$ and $\overline{\Lambda}$ particles are calculated with the account for feed-down effects due to the weak, $\Xi\to \Lambda +\pi$, and electromagnetic, $\Sigma^0\to \Lambda + \pi^0$, decays. The feed-down reduces strongly the particle polarization. Nevertheless, we can quantitatively describe the available STAR data for $\Lambda$ and $\overline{\Lambda}$ polarizations at $\sqrt{s_{NN}}=7.7$ and 11.5\,GeV, except the strong $\overline{\Lambda}$ polarization signal at 7.7\,GeV, which remained underpredicted. We found the (anti-)hyperon polarization, first, almost linearly increases with the collision centrality, but then decreases for peripheral collisions with centralities $\gsim 70\%$. This finding follows the general tendency seen in the available experimental data.
		
		\vspace{6pt}
		
		
		
		\authorcontributions{Authors N.Ts., E.K., V.V  equally contributed to conceptualization, formal analysis, and writing---review and editing. All authors have read and agreed to the published version of the manuscript.}
		
		\funding{This research was partially supported by the Slovak grant VEGA~1/0353/22.
			The calculations were performed on the ``Govorun'' computational cluster provided by the Laboratory of Information Technologies of JINR, Dubna.
		}
		
		\institutionalreview{Not applicable.}
		
		\informedconsent{Not applicable.}
		
		\dataavailability{No explicit data set are prepared.}
		
		\acknowledgments{We thank D.N.~Voskresensky and Yu.B.~Ivanov for discussions.}
		
		\conflictsofinterest{The authors declare no conflict of interest.}
		
		
		
		\abbreviations{Abbreviations}{
			The following abbreviations are used in this manuscript:\\
			
			\noindent
			\begin{tabular}{@{}ll}
				EoS & Equation of State\\
				HIC & Heavy-Ion Collision\\
				PHSD & Parton Hadron String Dynamics\\
				TLI & Time of the Last Interaction\\
				UrQMD & Ultra-relativistic Quantum Molecular Dynamics
			\end{tabular}
		}
		
		\begin{adjustwidth}{-\extralength}{0cm}
			
			\reftitle{References}

			\PublishersNote{}
		\end{adjustwidth}

\begin{thebibliography}{999}
				
				
				\bibitem{Adamczyk-Nature} Adamczyk, L.; Adkins, J.K.; Agakishiev, G.; Aggarwal, M.M.; Ahammed, Z.; Ajitanand, N.N.; Alekseev, I.; Anderson, D.M.; Aoyama, R.; Aparin, A.; et al. Global $\Lambda$ hyperon polarization in nuclear collisions.
				\emph{Nature} \textbf{2017}, \emph{548}, 62.
				
				\bibitem{Harris-Lb}
				Harris, J.W.; Sandoval, A.; Stock, R.; Stroebele, H.; Renfordt, R.E.; Geaga, J.V.; Pugh, H.G.; Schroeder, L.S.; Wolf, K.L.; Dacal, A. $\Lambda$ production near threshold in central nucleus-nucleus collisions. \emph{Phys. Rev. Lett.} \textbf{1981}, \emph{47}, 229.
				
				\bibitem{Anikina} Anikina, M.; Aksinenko, V.; Dementiev, E.; Gaździcki, M.; Glagoleva, N.; Golokhvastov, A.; Goncharova, L.; Grachov, A.; Iovchev, K.; Kadykova, S.; et al.
				Characteristics of $\Lambda$ and K$^0$ particles produced in central nucleus-nucleus Collisions at a 4.5\,GeV/$c$ momentum per incident nucleon. \emph{Z. Phys. C} \textbf{1984}, \emph{25}, 1.
				
				\bibitem{Panagiotou-86}
				Panagiotou, A.P. $\Lambda$ nonpolarization: possible signature of quark matter. \emph{Phys. Rev. C} \textbf{1986}, \emph{33}, 1999.
				
				\bibitem{DeGrand-81}
				De Grand, T.; Miettinen, H. Quark dynamics of polarization in inclusive hadron production. \emph{Phys. Rev. D} \textbf{1981}, \emph{23}, 1227.
				
				\bibitem{Becattini-Lisa-2020}
				Becattini, F.; Lisa, M.A. Polarization and vorticity in the quark–gluon plasma. \emph{Annu. Rev. Nucl. Part. Sci.} \textbf{2020}, \emph{70}, 395.
				
				\bibitem{Becattini-Tinti2010}
				Becattini, F.; Tinti, L. The ideal relativistic rotating gas as a perfect fluid with spin. \emph{Ann. Phys. (NY)} \textbf{2010}, \emph{325}, 1566.
				
				\bibitem{Becattini-Chandra2013}
				Becattini, F.; Chandra, V.; Del Zanna, L.; Grossi, E. Relativistic distribution function for particles	with spin at local thermodynamical equilibrium.
				\emph{Ann. Phys. (NY)} \textbf{2013}, \emph{338}, 32.
				
				\bibitem{Fang-Pang-Wang2016}
				Fang, R.-h.; Pang, L.-g.; Wang, Q.; Wang, X.-n. Polarization of massive fermions in a vortical fluid. \emph{Phys. Rev. C} \textbf{2016}, \emph{94}, 024904.
				
				\bibitem{Becattini-Karpenko-Lisa2017}
				Becattini, F.; Karpenko, I.; Lisa, M.A.; Upsal, I.; Voloshin, S.A. Global hyperon polarization at local thermodynamic equilibrium with vorticity, magnetic field, and feed-down. 
				\emph{Phys. Rev. C} \textbf{2017}, \emph{95}, 054902.
				
				\bibitem{Karpenko-Becattini2017}
				Karpenko, I.; Becattini, F. Study of $\Lambda$ polarization in relativistic nuclear collisions at $\sqrt{s_{NN}} = 7.7$--200\,GeV,
				\emph{Eur. Phys. J. C} \textbf{2017}, \emph{77}, 213.
				
				\bibitem{Xie-Wang-Csernai2017}
				Xie, Y.; Wang, D.; Csernai, L.P. Global $\Lambda$ polarization in high energy collisions. \emph{Phys. Rev. C} \textbf{2017},  \emph{95}, 031901(R).
				
				\bibitem{Ivanov-PRC100}
				Ivanov, Y.B.; Toneev, V.D.; Soldatov, A.A. Estimates of hyperon polarization in heavy-ion collisions at collision energies
				$\sqrt{s_{NN}} = 4$--40\,GeV. \emph{Phys. Rev. C} \textbf{2019}, \emph{100}, 014908
				
				\bibitem{Ivanov-PRC102}
				Ivanov, Y.B.; Soldatov, A.A. Correlation between global polarization, angular momentum, and flow in heavy-ion collisions. \emph{Phys. Rev. C} \textbf{2020}, \emph{102}, 024916.
				
				
				\bibitem{Ivanov-PRC103}
				Ivanov, Y.B. Global $\Lambda$ polarization in moderately relativistic nuclear collisions. \emph{Phys. Rev. C} \textbf{2021}, {\emph{103}}, L031903.
				
				\bibitem{Ivanov-PRC105}
				Ivanov, Y.B.; Soldatov, A.A. Global $\Lambda$ polarization in heavy-ion collisions at energies 2.4--7.7 GeV: Effect of meson-field interaction.
				\emph{Phys. Rev. C} \textbf{2022}, \emph{105}, 034915.
				
				\bibitem{Li-Pang-Wang-Xia-PRC96}
				Li, H.; Pang, L.G.; Wang, Q.; Xia, X.L. Global $\Lambda$ polarization in heavy-ion collisions from a transport model. \emph{Phys. Rev. C} \textbf{2017}, \emph{96}, 054908. 
				
				\bibitem{Sun-Ko-PRC96}
				Sun, Y.; Ko, C.M. $\Lambda$ hyperon polarization in relativistic heavy ion collisions from a chiral kinetic approach. \emph{Phys. Rev.C} \textbf{2017}, \emph{96}, 024906. 
				
				\bibitem{KTV-PRC97}
				Kolomeitsev, E.E.; Toneev, V.D.; Voronyuk, V. Vorticity and hyperon polarization at energies available at JINR Nuclotron-based Ion Collider fAcility. \emph{Phys. Rev. C} \textbf{2018}, \emph{97}, 064902. 
				
				\bibitem{Wei-Deng-Huang-PRC99}
				Wei, D.X.; Deng, W.T.; Huang, X.G. Thermal vorticity and spin polarization in heavy-ion collisions. \emph{Phys. Rev. C} \textbf{2019}, \emph{99}, 014905.
				
				\bibitem{Shi-Li-Liao-PLB788}
				Shi, S.; Li, K.; Liao, J. Searching for the subatomic swirls in the CuCu and CuAu collisions. \emph{Phys. Lett. B} \textbf{2019}, \emph{788}, 409. 
				
				\bibitem{Vitiuk-BZ2020}
				Vitiuk, O.; Bravina, L.V.; Zabrodin, E.E. Is different $\Lambda$ and $\overline{\Lambda}$ polarization caused by different spatio-temporal freeze-out picture? \emph{Phys. Lett. B} \textbf{2020}, \emph{803}, 135298. 
				
				\bibitem{Yassine:2022}
				Yassine, R.A.; Adamczewski-Musch, J.; Asal, C.; Becker, M.; Belounnas, A.; Blanco, A.; Blume, C.; Chlad, L.; Chudoba, P.; Ciepal, I.; Cordts, M.; et al. Measurement of global polarization of $\Lambda$ hyperons in few-GeV heavy-ion collisions. \emph{Phys. Lett. B} \textbf{2022}, {\emph {835}}, 137506.
				
				\bibitem{Ivanov-PRC102-AVE}
				Ivanov, Y.B. Global polarization in heavy-ion collisions based on the axial vortical effect. \emph{Phys. Rev. C} \textbf{2020}, \emph{102}, 044904.
				
				
				
				\bibitem{helicity:2022}
				Tsegelnik, N.S.; Kolomeitsev, E.E.; Voronyuk, V., Helicity and vorticity in heavy-ion collisions at NICA energies. {\emph{arXiv}} \textbf{2022}, {arXiv:2211.09219}.
				
				\bibitem{PHSD}
				Bratkovskaya, E.L.; Cassing, W.; Konchakovski, V.P.; Linnyk, O. Parton-Hadron-String Dynamics at Relativistic Collider Energies.
				\emph{Nucl. Phys. A} \textbf{2011}, \emph{856}, 162.
				\bibitem{PHSD-contin}
				Cassing, W.; Tolos, L.; Bratkovskaya, E.L.; Ramos, A. Anti-kaon production in A+A collisions at SIS energies within an off-shell G matrix approach. \emph{Nucl. Phys. A} \textbf{2003}, \emph{727}, 59.
				\bibitem{Birdsall1997}
				Birdsall, C.K.; Fuss, D. Clouds-in-Clouds, Clouds-in-Cells Physics for Many-Body Plasma Simulation. \emph{J. Comp. Phys.} \textbf{1997}, \emph{135},~141.
				
				\bibitem{SDM09}
				Satarov, L.M.; Dmitriev, M.N.; Mishustin, I.N.
				Equation of state of hadron resonance gas and the phase diagram of strongly interacting matter.
				\emph{Phys. Atom. Nucl.} \textbf{2009}, \emph{72}, 1390.
				
				\bibitem{KT-HysHSD}
				Khvorostukhin, A.S.; Toneev, V.D. Rapidity distributions of hadrons in the HydHSD hybrid model. \emph{Phys. Atom. Nucl.} \textbf{2017}, \emph{80},~285.
				
				\bibitem{KKT-Hydro}
				Khvorostukhin, A.S.; Kolomeitsev, E.E.; Toneev, V.D. Hybrid model with viscous relativistic hydrodynamics: A role of constraints on the
				shear-stress tensor. \emph{Eur. Phys. J. A} \textbf{2021}, \emph{57}, 294.
				\bibitem{Ivanov-rings}
				Ivanov, Y.; Soldatov, A.A. Vortex rings in fragmentation regions in heavy-ion collisions at $\sqrt{{s}_{NN}}=39$ GeV. \emph{Phys. Rev. C} \textbf{2018}, \emph{97}, 044915.
				
				\bibitem{Ivanov-rings-new}
				Ivanov, Y.B. Vortex rings in heavy-ion collisions at energies $\sqrt{s_{NN}}=$ 3--30 GeV and possibility of their observation. \emph{arXiv} \textbf{2022}, {arXiv:2211.17190}.
				
				\bibitem{BGST-Hseparation}
				Baznat, M.I.; Gudima, K.K.; Sorin, A.S.; Teryaev, O.V. Helicity separation in heavy-ion collisions. \emph{Phys. Rev. C} \textbf{2013}, {\emph{88}}, 061901(R).
				\bibitem{BGST-Vsheet}
				Baznat, M.I.; Gudima, K.K.; Sorin, A.S.; Teryaev, O.V. Femto-vortex sheets and hyperon polarization in heavy-ion collisions. \emph{Phys. Rev. C} \textbf{2016}, \emph{93}, 031902.
				\bibitem{HSD}
				Cassing, W.; Bratkovskaya, E.L. Hadronic and electromagnetic probes of hot
				and dense nuclear matter. \emph{Phys. Rep.} \textbf{1999}, {\emph{308}}, 65.
				\bibitem{Cassing:2015owa}
				Cassing, W.; Palmese, A.; Moreau, P.; Bratkovskaya, E.L. Chiral symmetry restoration versus deconfinement in heavy-ion collisions at high baryon density. \emph{Phys. Rev. C} \textbf{2016}, {\emph{93}}, 014902.
				
				
				\bibitem{NILSSONALMQVIST1987387}
				Bo, N.-A.; Evert, S. Interactions between hadrons and nuclei: The lund monte carlo - fritiof version 1.6 -.
				\emph{Comp. Phys. Comm.} \textbf{1987}, \emph{43}, 387.
				\bibitem{Andersson:1992iq}
				Andersson, B.; Gustafson, G.; Pi, H. The FRITIOF model for very high-energy hadronic collisions.
				\emph{Z. Phys. C} \textbf{1993}, \emph{57}, 485.
				\bibitem{Schwinger}
				Schwinger, J. On gauge invariance and vacuum polarization. \emph{Phys. Rev.} \textbf{1951}, {\emph{82}}, 664.
				
				\bibitem{GCG-1998}
				Geiss, J.; Cassing, W.; Greiner, C. Strangeness production in the hsd transport approach from SIS to SPS
				energies. \emph{Nucl. Phys. A} \textbf{1998}, {\emph{644}}, 107.
				\bibitem{Palmese:2016rtq}
				Palmese, A.; Cassing, W.; Seifert, E.; Steinert, T.; Moreau, P.; Bratkovskaya, E.L. Chiral symmetry restoration in heavy-ion collisions at intermediate energies. \emph{Phys. Rev. C} \textbf{2016}, \emph{94}, 044912.
				
				\bibitem{Adam:2019koz}
				Adam, J.; Adamczyk, L.; Adams, J.R.; Adkins, J.K.; Agakishiev, G.; Aggarwal, M.M.; Ahammed, Z.; Alekseev, I.; Anderson, D.M.; Aoyama, R.; et~al. Strange hadron production in Au+Au collisions at $ \sqrt{s_{NN}} = $7.7 , 11.5, 19.6, 27, and 39\,GeV. 
				\emph{Phys. Rev. C} \textbf{2020}, \emph{102}, 034909.
				\bibitem{Becattini:2015ska}
				Becattini, F.; Inghirami, G.; Rolando, V.; Beraudo, A.; Del Zanna, L.; De Pace, A.; Nardi, M.; Pagliara, G.; Chandra, V. A study of vorticity formation in high energy nuclear collisions. \emph{Eur. Phys. J. C} \textbf{2015}, {\emph{75}}, 406.
				\bibitem{PDG:2020}
				Zyla, P.; Barnett, R.M.; Beringer, J.; Dahl, O.; Dwyer, D.A.; Groom, D.E.; Lin, C.J.; Lugovsky, K.S.; Pianori, E.; Robinson, D.J.; et al. Review of Particle Physics. \emph{PTEP} \textbf{2020},
				\emph{2020}, 083C01.
				
				\bibitem{Okubo:2022}
				Okubo, K.; STAR Collaboration. Measurement of global polarization of $\Lambda$ hyperons in Au+Au $\sqrt{s_{NN}} = 7.2$\,GeV fixed target collisions at RHIC-STAR experiment. \emph{EPJ Web Conf.} \textbf{2022}, {\emph{259}}, 06003.
				
				\bibitem{STAR:2021beb}
				Abdallah, M.S.; Aboona, B.E.; Adam, J.; Adamczyk, L.; Adams, J.R.; Adkins, J.K.; Agakishiev, G.; Aggarwal, I.; Aggarwal, M.M.; Ahammed, Z.; et al. Global $\Lambda$-hyperon polarization in Au+Au collisions at $\sqrt {s_{NN}}$=3~GeV. 
				\emph{Phys. Rev. C }\textbf{2021}, {\emph{104}}, L061901.
				\bibitem{Adam:2018}
				Adam, J.; Adamczyk, L.; Adams, J.R.; Adkins, J.K.; Agakishiev, G.; Aggarwal, M.M.; Ahammed, Z.; Ajitanand, N.N.; Alekseev, I.; Anderson, D.M.; et al. Global polarization of $\Lambda$ hyperons in Au + Au collisions at $\sqrt{{s}_{NN}}=200$ GeV. \emph{Phys. Rev. C} \textbf{2018}, {\emph{98}}, 014910.
				
				
				
				
				
				
				
				
				
				\bibitem{Liang_2021}
				Liang, Z.-T.; Song, J.; Upsal, I.; Wang, Q.; Xu, Z., Rapidity dependence of global polarization in heavy ion collisions. \emph{Chin. Phys. C} \textbf{2021}, {\emph{45}}, 014102.
				\bibitem{Ayala:2022yyx}
				Ayala, A.; Dom\'\i{}nguez, I.; Maldonado, I.; Tejeda-Yeomans, M.E. $\Lambda$ and $\overline{\Lambda}$ global polarization from the core-corona model. \emph{Rev. Mex. Fis. Suppl.} \textbf{2022}, {\emph{3}}, 040914.
				
				
				\bibitem{Alzhrani:2022dpi}
				Alzhrani, S.; Ryu, S.; Shen, C. \ensuremath{\Lambda} spin polarization in event-by-event relativistic heavy-ion collisions. \emph{Phys. Rev. C} \textbf{2022}, {\emph{106}}, 014905.
				
				
			\end{thebibliography}
	\end{document}